\documentclass[pre,reprint,eqsecnum,superscriptaddress,twocolumn,longbibliography,floatfix]{revtex4-2}

\usepackage{empheq}
\usepackage{graphicx}
\usepackage{dcolumn}
\usepackage{bm}

\usepackage[utf8]{inputenc}
\usepackage[T1]{fontenc}
\usepackage{mathptmx}
\usepackage{etoolbox}
\usepackage{xcolor}
\usepackage{enumitem}
\usepackage{mathpazo}
\usepackage{bm}
\usepackage[unicode=true,
 bookmarks=true,bookmarksnumbered=false,bookmarksopen=false,
 breaklinks=false,pdfborder={0 0 0},pdfborderstyle={},backref=false,colorlinks=true]
 {hyperref}
\hypersetup{pdfborderstyle={},pdfborderstyle={},pdfborderstyle={},pdfborderstyle={},pdfborderstyle={},pdfborderstyle={},linkcolor=blue,citecolor=blue}
\usepackage{breakurl}

\usepackage{natbib}
\usepackage{amssymb,amsmath}

\newcommand{\bp}{\beta p}
\newcommand{\ehalf}{\frac{\epsilon}{2}}
\newcommand{\clp}{\text{cp}}
\newcommand{\beq}{\begin{equation}}
\newcommand{\eeq}{\end{equation}}
\newcommand{\beqa}{\begin{eqnarray}}
\newcommand{\eeqa}{\end{eqnarray}}
\newcommand{\nn}{\nonumber\\}
\def\bal#1\eal{\begin{align}#1\end{align}}

\makeatletter
\def\@email#1#2{%
 \endgroup
 \patchcmd{\titleblock@produce}
  {\frontmatter@RRAPformat}
  {\frontmatter@RRAPformat{\produce@RRAP{*#1\href{mailto:#2}{#2}}}\frontmatter@RRAPformat}
  {}{}
}%
\makeatother
\begin{document}

\title[Equilibrium properties of SW and SS disks in single-file confinement]{Exact equilibrium properties of square-well and square-shoulder disks in single-file confinement }
\author{Ana M. Montero}
\affiliation{Departamento de F\'isica, Universidad de Extremadura, E-06006 Badajoz, Spain}
\email{anamontero@unex.es}
\author{Andr\'es Santos}%
\affiliation{Departamento de F\'isica, Universidad de Extremadura, E-06006 Badajoz, Spain}
 \affiliation{
Instituto de Computaci\'on Cient\'ifica Avanzada (ICCAEx), Universidad de Extremadura, E-06006 Badajoz, Spain
}%
\email{andres@unex.es}

\date{\today}

\begin{abstract}
This study investigates the (longitudinal) thermodynamic and structural characteristics of single-file confined square-well and square-shoulder disks by employing a mapping technique that transforms the original system into a one-dimensional polydisperse mixture of  nonadditive rods. Leveraging standard statistical-mechanical techniques, exact results are derived for key properties, including the equation of state, internal energy, radial distribution function, and structure factor. The asymptotic behavior of the radial distribution function is explored, revealing structural changes in the spatial correlations. Additionally, exact analytical expressions for the second virial coefficient are presented.
The theoretical results for the thermodynamic and structural properties are validated by our Monte Carlo simulations.
\end{abstract}

\maketitle

\section{\label{sec:introduction}Introduction}

The study of the thermodynamic and structural properties of liquids whose particles interact via simple potentials has been a field of interest for many years~\cite{BH76,HM13,EL85,BCT07,HS18,PMPSGLVTC22, MCSM22,LK78}. In this context, ``simple'' refers to pairwise potentials that are relatively straightforward and uncomplicated in form and mathematical representation, involving basic functional forms. The primary rationale behind this focus is to enable a profound understanding of system behavior while retaining key realistic features similar to those observed in actual fluids.

Within the realm of these elementary potentials, two that stand out prominently are the square-well (SW)~\cite{BH67, BH67b, LK78,YS94,MRR09} and square-shoulder (SS)~\cite{LKLLW99,YSH11,BOO13,GSC11} potentials. They are characterized by an impenetrable hard core paired with either an attractive well or a repulsive step. The SS potential is purely repulsive and belongs to the family of the so-called core-softened potentials, which have been widely used to study metallic liquids~\cite{SY76} or water anomalies~\cite{J04,BSB11,HU13,HU14,SSGBS01}. Conversely, the SW potential comprises a repulsive hard core complemented by an attractive well, making it suitable for modeling more intricate fluids governed by competing interactions~\cite{AIPR08,PMPSGLVTC22}.

Although bulk fluids of particles interacting with these two potentials have been thoroughly studied using different approaches, to the best of our knowledge, little is still known about their behavior in confined geometries~\cite{RPSP10,F23}.
Confined liquids manifest in diverse scenarios, spanning from biological systems to material science. Unraveling the distinctions in their properties compared to bulk liquids constitutes a pivotal stride toward comprehending their behavior in entirety~\cite{RGM16,Ketal11,FLS12,*FLS13,*FS22,JF22}.

This paper focuses on highly confined SW and SS two-dimensional (2D) systems, where the length of one of the dimensions is much larger than that of the other one, the latter being so small as to confine particles into single-file formation. In such a scenario, the system can be treated as quasi one-dimensional (q1D)~\cite{B62,B64b,WPM82,PK92,KP93,P02,KMP04,FMP04,VBG11,GV13,GM14,M14b,*M15,GM15,HFC18,M20,HBPT20,P20,MBGM22,MS23,MS23b,F23,PBT23}, and
its most relevant properties are the longitudinal ones.

Our study is motivated by experiments on confined q1D colloidal liquids, which have revealed an attractive potential well within the effective colloid-colloid interactions \cite{CLSR02}. Additionally, it is well established that effective electrostatic interactions between colloids in colloid-nanoparticle mixtures can be modeled with a hard-core plus a repulsive potential \cite{D17}.

In these circumstances, the advantage of using confined SW and SS disks over more complex potentials becomes clear.
The significance of confined systems with exact solutions is evident, as they not only facilitate a more profound exploration of their physical properties, but also serve as a reliable benchmark for assessing the accuracy of approximate methods and computer simulations. This, in turn, enhances their utility in studying more intricate systems~\cite{TL94a,TL94b,BE02, MS06}.

While adapting the standard transfer-matrix method (TMM) \cite{KP93} to SW and SS potentials allows for the derivation of thermodynamic quantities, obtaining structural properties with the TMM is much more challenging. Due to this, we employ an exact mapping technique that transforms the system into a one-dimensional (1D) polydisperse mixture of \emph{nonadditive} rods with equal chemical potential~\cite{MS23,MS23b}. This approach differs from the approximate mapping proposed by Post and Kofke \cite{PK92} for  the hard-disk and hard-sphere cases, where ``\ldots the collision diameter of each pair of rods is given by the arithmetic
mean of their molecular diameters.''

The structure of our paper is the following: Section~\ref{sec:system} describes the confined system, along with its main properties, and establishes the equivalence between the confined system and its 1D mixture counterpart. Section~\ref{sec:solution} presents the exact theoretical results for  its main (longitudinal) thermodynamic and structural properties and a derivation of the second virial coefficient and the Boyle temperature, while Sec.~\ref{sec:montecarlo} is devoted to a brief description of our own Monte Carlo (MC) simulations. In Sec.~\ref{sec:results}, an analysis of all results is presented, with information on the transverse density  profile, the equation of state, the internal energy, the radial distribution function, and the structure factor. Finally, some concluding remarks are provided in Sec.~\ref{sec:concluding}.

\section{\label{sec:system}The Confined SW and SS Fluids}

\subsection{\label{sec:solsystem}The 2D system}

\begin{figure}
	\includegraphics[width=\columnwidth]{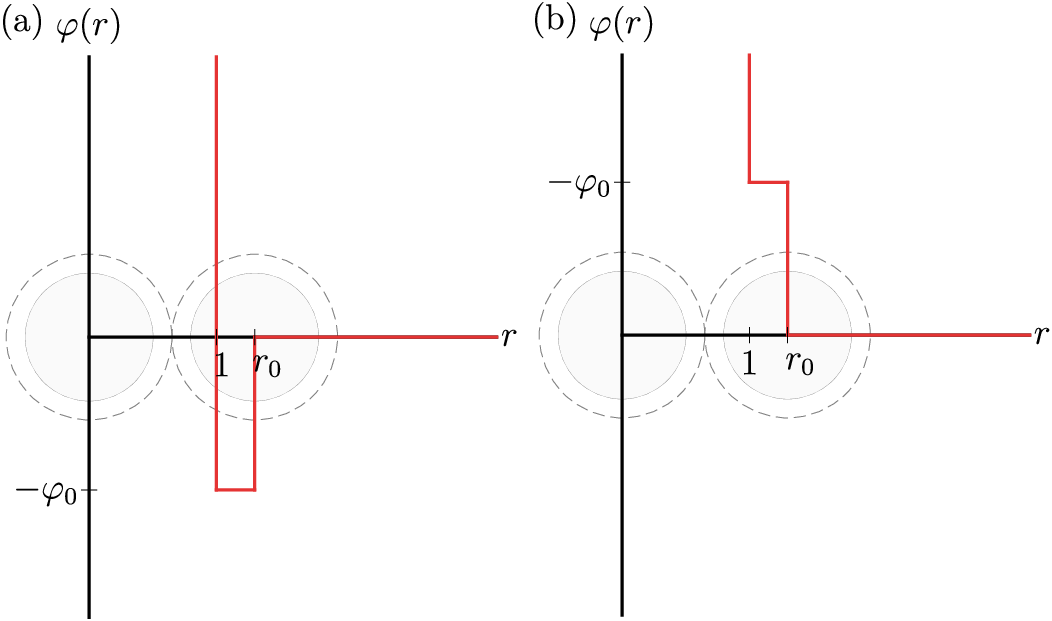}
	\caption{Schematic representation of (a) the SW potential and (b) the SS potential.}
	\label{fig:potential}
\end{figure}

We consider a 2D system of $N$ particles interacting via a pairwise potential,
\begin{equation}
\label{varphi(r)}
	\varphi(r) =
	\begin{cases}
		\infty & \text{if } r<1, \\
		-\varphi_0 & \text{if } 1<r<r_0, \\
		0 & \text{if } r>r_0, \\
	\end{cases}
\end{equation}
where $r_0$ is the range of interaction and, for simplicity, the hard-core diameter of the particles defines the unit of length. The sign of $\varphi_0$ determines whether, in addition to the hard core, the potential has an attractive corona ($\varphi_0>0$, SW) or a repulsive one ($\varphi_0<0$, SS). A schematic representation of both potentials is shown in Fig.~\ref{fig:potential}.
The depth of the well ($\varphi_0$) or the height of the shoulder ($-\varphi_0$) allows us to define a reduced temperature $T^*=k_BT/|\varphi_0|$, where $T$ is the absolute temperature and $k_B$ is the Boltzmann constant.
This ensures that $T^*$ is always  positive. An alternative definition, $T^* = k_B T / \varphi_0$, would result in negative values in the SS case, which could be confusing.

\begin{figure}
	\includegraphics[width=\columnwidth]{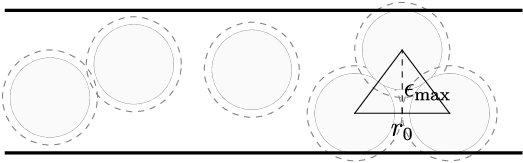}
	\caption{Schematic representation of the particles confined in a narrow channel}
	\label{fig:system}
\end{figure}

The particles are assumed to be confined in a very long rectangular channel of width $w=1+\epsilon$, where the excess pore width ($\epsilon$) is the available space for the particle centers, and length $L\gg w$.  To avoid second-nearest neighbor interactions, for any given value of the corona diameter ($r_0$), the maximum value of the excess pore width is limited to $\epsilon_{\mathrm{max}}=\sqrt{1-r_0^2/4}$, as shown in Fig.~\ref{fig:system}. Under these conditions, the channel is narrow enough to prevent the particles from bypassing each other, forcing them into a single file. Note also that the particles interact with the walls only through the hard core diameter.

In general, if two particles $\alpha$ and $\alpha'$ are in close contact (i.e., $r_{\alpha\alpha'}=1$) with a transverse separation $|y_\alpha-y_{\alpha'}|$ between their centers,   their longitudinal separation  is $|x_{\alpha}-x_{\alpha'}|=a(y_{\alpha}-y_{\alpha'})$, where
\beq
\label{a(s)}
a(\Delta y)\equiv\sqrt{1-\Delta y^2}.
\eeq
Similarly, if the coronas of two particles are in contact (i.e., $r_{\alpha\alpha'}=r_0$), then $|x_{\alpha}-x_{\alpha'}|=b(y_{\alpha}-y_{\alpha'})$, where
\beq
\label{b(s)}
b(\Delta y)\equiv\sqrt{r_0^2-\Delta y^2}.
\eeq

Due to the high anisotropy between the transverse and longitudinal directions of this system, it is often useful to focus on its longitudinal properties, such as the number of particles per unit length, $\lambda \equiv N/L$ \footnote{The linear density $\lambda$ should be distinguished from the number of particles per unit area (areal density) $\rho\equiv N/L\epsilon=\lambda/\epsilon$.}, the longitudinal pressure $P_{\|}$ and the reduced pressure $p=\epsilon P_{\|}$. Note that there exists a close-packing density, $\lambda_{\text{cp}} = 1/a(\epsilon)$, at which pressure diverges.

For a given corona diameter $r_0$, the control parameters can be chosen as the excess pore width $\epsilon$, the reduced temperature $T^*$, and the linear density $\lambda$ (or, equivalently, the product $\bp$, where $\beta\equiv 1/k_BT$). In the high-temperature limit ($T^*\to\infty$), the attractive or repulsive corona becomes irrelevant and thus the system reduces to a pure hard-disk (HD) fluid, which has been well studied~\cite{B62,B64b,WPM82,PK92,KP93,P02,KMP04,FMP04,VBG11,GV13,GM14,M14b,*M15,GM15,HFC18,M20,HBPT20,P20,JF22,MBGM22,MS23,MS23b}. To make this property more explicit, suppose that $X$ is a quantity of dimensions (length)$^m$; then,
\bal
\label{T_infinity}
\lim_{T^*\to\infty}X^{\text{SW}}( \lambda,T^*;r_0,\epsilon)=&\lim_{T^*\to\infty}X^{\text{SS}}( \lambda,T^*;r_0,\epsilon)\nn
=&X^{\text{HD}}( \lambda;\epsilon).
\eal
In the opposite low-temperature limit ($T^*\to 0$), the SS particles become equivalent to HDs of diameter $r_0$; therefore,
\beq
\label{T_zero}
\lim_{T^*\to 0}X^{\text{SS}}( \lambda,T^*;r_0,\epsilon)=r_0^m X^{\text{HD}}( \lambda r_0;\epsilon/r_0).
\eeq

\subsection{\label{sec:solequiv}Equivalent 1D system}

In Appendix A, we argue that the properties of the confined 2D system described in Sec.~\ref{sec:introduction} can be exactly matched to those of an equivalent 1D polydisperse mixture, where the transverse coordinate of each particle,$-\epsilon/2<y<\epsilon/2$, plays the role of the dispersity parameter, and where the chemical potential of all components of the mixture is the same. While the original application of this equivalence was in the context of a HD fluid~\cite{MS23,MS23b}, it can be readily extended to any interaction potential $\varphi(r)$, with the caveat that interactions are constrained to nearest neighbors.

Although the equivalence holds precisely when the 1D mixture features a continuous distribution of components, practical considerations often demand the discretization of the system for numerical computations. Therefore, it usually proves more pragmatic to examine a 1D mixture with a \emph{discrete} but adequately large number of components, $M$, to accurately characterize the system. The theoretical expressions valid for the original continuous case can then be derived by considering the limit  $M\to\infty$.

In this discrete $M$-component mixture, each 1D component, indexed as $i=1,2,\ldots,M$, corresponds to a mapping of 2D particles with a transverse coordinate
\begin{equation}
\label{y_i}
	y_i=-\frac{\epsilon}{2}+(i-1)\delta y,\quad \delta y\equiv \frac{\epsilon}{M-1}.
\end{equation}
In turn, the 2D interaction potential $\varphi(r)$ translates into the 1D potential
\bal
	\label{varphi_ij}
	\varphi_{ij}(x)=&\varphi\left(\sqrt{x^2+(y_i-y_j)^2}\right)\nn
 =& \begin{cases} \infty &\text{if } x<a_{ij}, \\
		-\varphi_0&\text{if } a_{ij}<x<b_{ij},\\
		0 &\text{if } x>b_{ij},\end{cases}
\eal
where
\begin{subequations}
	\beq
	a_{ij}\equiv a(y_i-y_j),\quad
	b_{ij}\equiv b(y_i-y_j).
	\eeq
\end{subequations}

Within this framework, one can precisely ascertain the properties of the 1D mixture and  directly map them back onto the original 2D system.

\section{\label{sec:solution}Exact solution}

Most of the properties of 1D mixtures are derived in the isothermal-isobaric ensemble and can be described through the Laplace transform of the Boltzmann factor~\cite{S16},
\begin{equation}
\label{Om0}
	\Omega_{ij}(s,\beta) = \int_0^\infty {d}x\, e^{-sx}e^{-\beta \varphi_{ij}(x)},
\end{equation}
which, in the case of the 1D mixture described by Eq.~\eqref{varphi_ij}, yields
\begin{equation}\label{eq:omega}
\Omega_{ij}(s,\beta)=\frac{e^{\beta^*}}{s}\left[e^{-a_{ij}s}-\left(1-e^{-\beta^*}\right)e^{-b_{ij}s}\right].
\end{equation}
Here, $\beta^* \equiv \beta \varphi_0$. Note that $\beta^* = 1/T^* > 0$ for SW \emph{but} $\beta^* = -1/T^* < 0$ for SS. This way, henceforth, all expressions involving $\beta^*$ apply equally to both SW and SS cases.

In the standard theory of liquid mixtures, mole fractions are considered pre-determined thermodynamic variables. Yet, in the 1D mixture under consideration, the requirement for identical chemical potentials imposes specific conditions on the values of the mole fractions for each component. Let $\phi_i^2$ denote the mole fraction of component $i$. Then, the set $\{\phi_i\}$ is obtained by solving the eigenvalue equation
\begin{equation}\label{eq:eigenproblem}
	\sum_j\Omega_{ij}(\beta p,\beta)\phi_j=\frac{1}{ A^2}\phi_i,
\end{equation}
where $A$ is a quantity directly related to the chemical potential as $\beta \mu = \ln (A^2 \Lambda_{\mathrm{dB}})$, $ \Lambda_{\mathrm{dB}}=\sqrt{\beta/2\pi m}h$ being the thermal de Broglie wavelength.

While Eqs.~\eqref{Om0} and \eqref{eq:eigenproblem} emerge autonomously from the polydisperse 1D framework \cite{MS23b}, they turn out to coincide with the results one would obtain by applying the TMM. In the latter context, the Laplace transform of the Boltzmann factor evaluated at $s=\bp$, $\Omega_{ij}(\bp,\beta)$, is not but the $ij$ element of the transfer matrix.

\subsection{\label{sec:properties}Thermodynamic properties}

Two of the paramount thermodynamic quantities essential for computation in any equilibrium system are the equation of state and the excess internal energy per particle. The equation of state establishes a connection between pressure, density, and temperature, while the excess internal energy per particle encompasses the potential energy per particle (which, combined with the ideal-gas kinetic energy  $u_{\mathrm{id}}=\frac{1}{2}k_BT$, contributes to the overall internal energy per particle).

In general terms, the compressibility factor, $Z\equiv \beta p / \lambda $, and the excess internal energy per particle, $u_{\mathrm{ex}}$, of any given 1D mixture with equal chemical potentials are given by \cite{S16}
\begin{subequations}
\label{eos&uex}
\begin{equation}\label{eq:eos}
Z=-A^2\bp\sum_{i,j}\phi_i\phi_j\left[\frac{\partial\Omega_{ij}(\beta p,\beta)}{\partial \bp}\right]_{\beta},
\end{equation}
\begin{equation}\label{eq:uex}
u_{\mathrm{ex}}=-A^2\sum_{i,j}\phi_i\phi_j \left[\frac{\partial \Omega_{ij}(\beta p,\beta)}{\partial\beta}\right]_{\beta p}.
\end{equation}
\end{subequations}
Using Eq.~\eqref{eq:omega}, Eqs.~\eqref{eos&uex} become
\begin{subequations}
\label{eos&uex2}
\begin{equation}\label{eq:eos2}
	Z=1+A^2e^{\beta^*}\sum_{i,j}\phi_i\phi_j\left[a_{ij}e^{- \beta p a_{ij}}-b_{ij}\left(1-e^{-\beta^*}\right)e^{-\beta p b_{ij}}\right],
\end{equation}
\beq
\label{eq:uex2}
	\frac{u_{\text{ex}}}{\varphi_0}=-1+\frac{A^2}{\beta p}\sum_{i,j}\phi_i\phi_je^{-\beta p b_{ij}},
\eeq
\end{subequations}
where  we have used Eq.~\eqref{eq:eigenproblem} and the normalization condition $\sum_i\phi_i^2=1$.

\subsection{\label{sec:properties_g}Structural properties}

Contrary to thermodynamic properties, which relate to the global quantities of the system, structural properties are primarily concerned with the arrangements and configurations of the particles. The key advantage of the 1D mapping over the TMM lies precisely in its ability to access these structural properties. The fundamental structural property that can be examined is the (longitudinal) radial distribution function (RDF)
$g_{ij}(x)$, which, in Laplace space, is given by~\cite{MS23b}
\begin{align}\label{eq:rdf}
	\widetilde{G}_{ij}(s)=&\int_0^\infty dx\, e^{-sx} g_{ij}(x)\nn
	=&\frac{A^2}{\lambda \phi_i\phi_j}\left[\mathsf{\Omega}(s+\beta p)\cdot\left[\mathsf{I}-A^2\mathsf{\Omega}(s+\beta p)\right]^{-1}\right]_{ij},
\end{align}
where $\mathsf{\Omega}$ is the $M\times M$ matrix of elements $\Omega_{ij}$ and $\mathsf{I}$ is the unit matrix.
Henceforth, for enhanced  clarity, we will omit the second argument ($\beta$) in $\Omega_{ij}$.

The RDF in real space can be obtained by performing the inverse Laplace transform on Eq.~\eqref{eq:rdf}. The structure of the analytical form of  $g_{ij}(x)$ is presented in Appendix~\ref{app:rdfrealspace}.
At a practical level, we have used Eq.~\eqref{3.5ee} for $x\leq 3a(\epsilon)$. For $x> 3a(\epsilon)$, however, we have found  preferable to invert $\widetilde{G}_{ij}(s)$ numerically \cite{EulerILT}. Once the partial RDFs $g_{ij}(x)$ are known, the total RDF is obtained as
\begin{equation}\label{eq:rdffull}
	{g}(x)=\sum_{i,j} \phi_i^2 \phi_j^2 {g}_{ij}(x).
\end{equation}

The structure factor is another pivotal quantity that, although conveying the same physical information as the RDF, can be experimentally accessed through diffraction experiments. The 1D structure factor is directly linked to the Fourier transform of the total correlation function $h(x)\equiv g(x)-1$,
\begin{equation}
	S(q) = 1 + 2 \lambda \int_0^{\infty}{d}x\cos(qx)h(x).
	\end{equation}
In our scheme, this is equivalent to
\begin{equation}
	S(q) = 1 + \lambda \left[\widetilde{G}(s)+\widetilde{G}(-s) \right]_{s=\imath q},
\end{equation}
where $\widetilde{G}(s)=\sum_{i,j} \phi_i^2 \phi_j^2\widetilde{G}_{ij}(s)$ and $\imath$ is the imaginary unit.

\subsection{Compressibility factor in terms of the RDF}
For an arbitrary (nearest-neighbor) interaction potential $\varphi_{ij}(x)$, the compressibility factor $Z$ is given by Eq.~\eqref{eq:eos}, while the RDF $g(x)$ is given by Eqs.~\eqref{eq:rdf} and \eqref{eq:rdffull}. In both cases one first needs to evaluate the Laplace transform $\Omega_{ij}(s)$. The interesting question is, can one express $Z$ directly in terms of density and integrals involving $g(x)$?
An affirmative response can be found in Appendix~\ref{app:rdffromgx}, with the outcome
\begin{equation}
\label{eq:Zfromgx}
Z=\frac{1-\lambda\int_{a(\epsilon)}^{r_0} dx\, g(x)}{1-\lambda\left[r_0-\lambda\int_{a(\epsilon)}^{r_0} dx\, (r_0-x) g(x)\right]}.
\end{equation}
Equation~\eqref{eq:Zfromgx}, which generalizes Eq.~(2.13) of Ref.~\cite{F23}, can be conveniently employed in NVT simulations.

\subsection{Continuous polydisperse mixture}
To take the continuum limit, let us define the transverse density profile of the original 2D system by $\phi^2(y_i)=\phi_i^2/\delta y$, as well as the parameter $\ell=(\bp/A^2)\delta y$. Also, Eq.~\eqref{eq:omega} can be  written as
\begin{equation}\label{eq:omega2}
\Omega(y,y';s)=\frac{e^{\beta^*}}{s}\left[e^{-a(y-y')s}-\left(1-e^{-\beta^*}\right)e^{-b(y-y')s}\right].
\end{equation}

Now, taking the limit $M\to\infty$ (which implies $\delta y\to 0$), Eqs.~\eqref{eq:eigenproblem} and \eqref{eos&uex2} become
\begin{subequations}
\begin{equation}
\label{eq:eigenproblem2}
	\int_\epsilon dy'\,\Omega(y,y';\beta p)\phi(y')=\frac{\ell}{\bp}\phi(y),
\end{equation}
\bal
\label{eq:eos22}
	Z=&1+\frac{\bp}{\ell}e^{\beta^*}	\int_{\epsilon}dy\int_{\epsilon}dy'\,\left[a(y-y')e^{- \beta p a(y-y')}\right.\nn
&\left.-b(y-y')\left(1-e^{-\beta^*}\right)e^{-\beta p b(y-y')}\right]\phi(y)\phi(y'),
\eal
\beq
\label{eq:uex22}
	\frac{u_{\text{ex}}}{\varphi_0}=-1+\frac{1}{\ell}\int_{\epsilon}dy\int_{\epsilon}dy'\,e^{-\beta p b(y-y')}\phi(y)\phi(y').
\eeq
\end{subequations}
Here, we have adopted the notation  convention $\int_{\epsilon}d y\, \cdots\equiv \int_{-\ehalf}^{\ehalf}dy\, \cdots$.

In what concerns the structural properties, let us first rewrite Eq.~\eqref{eq:rdf} in the equivalent form
\beq
\label{B1cc}
\frac{\phi_j\widetilde{G}_{ij}(s)}{A^2}=\frac{\Omega_{ij}(s+\bp)}{\lambda\phi_i}+\sum_k \phi_k\widetilde{G}_{ik}(s)\Omega_{kj}(s+\bp),
\eeq
and define $g(y_i,y_j;x)=g_{ij}(x)$ in real space and $\widetilde{G}(y_i,y_j;s)=\widetilde{G}_{ij}(s)$ in Laplace space. Then, in the limit $M\to\infty$ we get the following linear integral equation of the second kind,
\bal
\label{B1ccc}
\frac{\ell \phi(y')\widetilde{G}(y,y';s)}{\bp}=&\frac{\Omega(y,y';s+\bp)}{\lambda\phi(y)}+\int_{\epsilon}dy''\, \phi(y'')\nn
&\times\widetilde{G}(y,y'';s)\Omega(y'',y';s+\bp).
\eal
In turn, Eq.~\eqref{eq:rdffull} becomes
\beq
g(x)=\int_{\epsilon}dy\int_{\epsilon}dy'\,\phi^2(y)\phi^2(y')g(y,y';x).
\eeq

Note that Eq.~\eqref{eq:Zfromgx} is still applicable in the continuum limit.

It is noteworthy that, within the TMM framework, the physical $\ell$ in Eq.~\eqref{eq:eigenproblem2} is the \emph{largest} eigenvalue. The second largest eigenvalue (in absolute value), $\ell_1$, provides valuable insights into transverse correlations among $n$th neighbor particles \cite{VBG11}. Let us consider a reference particle $0$ with a transverse coordinate $y_0$. The transverse correlation function $\langle y_0 y_n\rangle$, where $y_n$ is the transverse coordinate of the $n$th neighbor, is expected to be negative (positive) for odd (even) $n$ and to asymptotically decay exponentially with $n$: $\langle y_0 y_n\rangle\sim (-1)^n e^{-n/\xi_\perp}$. Here, $\xi_\perp=1/\ln|\ell/\ell_1|$ is the transverse correlation degree \footnote{We prefer to use the term ``transverse correlation degree'' instead of ``transverse correlation length'' to emphasize its dimensionless nature.}, a dimensionless quantity measuring the number of neighbors after which transverse positions become uncorrelated. In the equivalent polydisperse 1D framework, $\xi_\perp$ quantifies the decay of correlations between the identities (or ``species'') of $n$th-neighbor particles.

\subsection{Asymptotic behavior of the RDF}
\label{sec3E}

The asymptotic behavior of $g(y,y';x)$ is related to the nonzero poles, $\{s_n\}$, of $\widetilde{G}(y,y';s)$ and their associated residues. In general,
\begin{subequations}
\begin{equation}
	g(y,y';x) = 1+\sum_{n=1}^{\infty}\mathcal{A}_n(y,y') e^{s_n x},
\end{equation}
\bal
	\mathcal{A}_{n}(y,y')\equiv&\, \mathrm{Res}[\widetilde{G}(y,y';s)]_{s_n}\nn
=& \left[\frac{\partial}{\partial s}\frac{1}{\widetilde{G}(y,y';s)}\right]_{s=s_n}^{-1}.
\eal
\end{subequations}
The asymptotic decay of the total correlation function $h(y,y';x)\equiv g(y,y';x)-1$ is then determined by either the nonzero real pole $s=-\kappa$ or the pair of conjugate poles $-\kappa \pm \imath \omega$ with the smallest value of $\kappa$. In this framework, $\xi=\kappa^{-1}$ represents the  correlation length, measuring the scale of decay of the correlation function $h(y, y'; x)$ \footnote{Note that, unlike the transverse correlation degree $\xi_\perp$, the correlation length $\xi$ has dimensions of length.}. If the dominant poles are complex, $\omega$ represents the asymptotic oscillation frequency and   one has
\begin{equation}\label{eq:poleimag}
	h(y,y';x)\approx 2 |\mathcal{A}(y,y')|e^{-\kappa  x} \cos(\omega x  + \delta),
\end{equation}
for asymptotically large values of $x$, where  $\mathcal{A}(y,y')=|\mathcal{A}(y,y')|e^{\pm \imath \delta}$ is the complex residue.
Equation~\eqref{eq:poleimag} describes an oscillatory decay of $h(y,y';x)$.
If, however, the dominant pole is real (i.e., $\omega=0$), then
\begin{equation}\label{eq:polereal}
	h(y,y';x) \approx \mathcal{A}(y,y') e^{-\kappa  x},
\end{equation}
where the residue $\mathcal{A}(y,y')$ is also a real number and therefore the asymptotic decay is purely monotonic.

\subsection{\label{sec:B2}Second virial coefficient}
In the low-density (or low-pressure) regime, the compressibility factor can be expressed as
\bal
\label{5.3d}
  Z=&1+B_2\lambda+\mathcal{O}(\lambda^2)\nn
=&1+B_2\bp+\mathcal{O}(\bp^2),
  \eal
where $B_2$ is the second virial coefficient.

In general, the behavior of $\Omega(y,y';s)$ for small $s$ is of the form
	\begin{equation}
\label{5.3c}
	\Omega(y,y';s)=s^{-1}+\Psi(y,y')+\mathcal{O}(s),
\end{equation}
where $\Psi(y,y')$ does not need to be specified at this stage.
By following steps analogous to those in Appendix~B of Ref.~\cite{MS23}, one can prove that the low-pressure solution to the eigenvalue problem in Eq.~\eqref{eq:eigenproblem2} is
\begin{subequations}
\begin{equation}
	\phi(y)=\frac{1}{\sqrt{\epsilon}}\left[1+\overline{\phi}_1(y)\beta p +\mathcal{O}(\beta p^2)\right],
\end{equation}
	\begin{equation}
	\ell=\epsilon\left[1-B_2\beta p +\mathcal{O}(\beta p^2)\right],
\end{equation}
\end{subequations}
where
\begin{subequations}
\beq
\label{5.6a}
\overline{\phi}_1(y)=B_2+\frac{1}{\epsilon}\int_{\epsilon} dy'\,\Psi(y, y'),
\eeq
\beq
\label{B2}
B_2=-\frac{1}{\epsilon^2}\int_{\epsilon} dy\int_{\epsilon} dy'\,\Psi(y, y').
\eeq
\end{subequations}

In the particular case of the SW or SS potentials, from Eq.~\eqref{eq:omega2} we can easily identify the function $\Psi(y,y')$ as
\beq
\Psi(y,y')=-e^{\beta^*}a(y-y')+\left(e^{\beta^*}-1\right)b(y-y').
\eeq
Insertion into Eq.~\eqref{B2} yields
\begin{equation}
B_2(T^*;r_0,\epsilon)=e^{\beta^*}B_2^{\text{HD}}(\epsilon)-\left(e^{\beta^*}-1\right)r_0 B_2^{\text{HD}}(\epsilon/r_0),
\end{equation}
where
\beq
\label{B2_SW&SS}
B_2^{\text{HD}}(\epsilon)=\frac{2}{3}\frac{ \left(1+\frac{\epsilon ^2}{2}\right) \sqrt{1-\epsilon ^2}-1}{\epsilon ^2}+\frac{\sin ^{-1}(\epsilon )}{\epsilon }
\eeq
is the second virial coefficient of the confined HD fluid.
As expected from Eqs.~\eqref{T_infinity} and \eqref{T_zero}, $\lim_{T^*\to\infty}B_2^{\text{SW}}(T^*;r_0,\epsilon)=\lim_{T^*\to\infty}B_2^{\text{SS}}(T^*;r_0,\epsilon)=B_2^{\text{HD}}(\epsilon)$ and $\lim_{T^*\to 0}B_2^{\text{SS}}(T^*;r_0,\epsilon)=r_0 B_2^{\text{HD}}(\epsilon/r_0)$.

In the SS case ($\beta^*<0$), the second virial coefficient is positive definite. However, in the SW case ($\beta^*>0$), it changes from negative to positive values as temperature increases; the temperature at which $B_2^{\text{SW}}=0$ defines the Boyle temperature
\beq
\label{TBoyle}
T^*_{\text{B}}=-\frac{1}{\ln\left[1-B_2^{\text{HD}}(\epsilon)/r_0 B_2^{\text{HD}}(\epsilon/r_0)\right]}.
\eeq
At fixed $r_0$, $T^*_{\text{B}}$ increases as $\epsilon$ increases from $\epsilon=0$ to $\epsilon=\epsilon_{\max}=\sqrt{1-r_0^2/4}$.

The thermodynamic Maxwell relation $\bp (\partial u/\partial\bp)_\beta=(\partial Z/\partial\beta)_{\bp}$  allows us to obtain $u_{\mathrm{ex}}/\varphi_0=(\partial B_2/\partial\beta^*)\bp +\mathcal{O}(\bp^2)$. From Eq.~\eqref{B2_SW&SS} we get $\partial B_2/\partial\beta^*=-e^{\beta^*}\left[r_0 B_2^{\text{HD}}(\epsilon/r_0)-B_2^{\text{HD}}(\epsilon)\right]$.

It is known that truncation of the virial series in powers of pressure is much more accurate than truncation of the series in powers of density \cite{MS23,MSRH11}. Thus, truncating at the order of the second virial coefficient in the expansion in powers of pressure yields the following approximate equations,
  \beq
  \label{Z&u_approx}
  Z\approx \frac{1}{1-B_2\lambda},\quad \frac{u_{\mathrm{ex}}}{\varphi_0}\approx \frac{(\partial B_2/\partial\beta^*)\lambda}{1-B_2\lambda}.
  \eeq

\section{\label{sec:montecarlo}Monte Carlo simulations}
To test the theoretical results presented in Sec.~\ref{sec:solution} for the thermodynamic properties (compressibility factor and internal energy), we have performed isothermal-isobaric (NPT) MC simulations on the original 2D confined system, in which the excess pore width $\epsilon$ and the longitudinal pressure $p$ are kept fixed but the longitudinal length $L$ fluctuates. For the investigation of structural properties, we found it more convenient to employ canonical (NVT) MC simulations.

We have  checked  the equivalence of results between the NVT and NPT ensembles for both thermodynamic and structural properties, as well as the consistency with the NVT MC data reported in Ref.~\cite{F23}. Whereas  NVT simulations do not provide direct access to the equation of state,  the compressibility factor can be computed from $g(x)$  through Eq.~\eqref{eq:Zfromgx}.
Nevertheless, from a practical point of view,  the values of $Z$ computed in this manner for large densities become extremely sensitive to numerical errors in the evaluation of the integrals $\int_{a(\epsilon)}^{r_0} dx\, g(x)$ and $\int_{a(\epsilon)}^{r_0} dx\, (r_0-x) g(x)$ \cite{F23}, which makes the NPT ensemble much more suitable to compute the equation of state.

In general, $10^7$ samples were collected from a system with $10^2$ particles, after an equilibration process of at least $10^7$ configurations and with an acceptance ratio of roughly $50\%$.

\section{\label{sec:results}Results}

As shown in Sec.~\ref{sec:solsystem}, a SW or SS interaction potential of range $r_0$ sets the maximum value of the excess pore width to $\epsilon_{\mathrm{max}}=\sqrt{1-r_0^2/4}$. The two limiting cases for these values correspond to the pure 1D system ($r_0=2, \, \epsilon=0$) and to the confined HD fluid ($r_0=1, \, \epsilon=\sqrt{3}/2\simeq 0.866$), both of them already studied exactly in the literature~\cite{FW69,KP93,S16,MS23,MS23b}.

As a compromise between introducing a nonnegligible corona and, at the same time, departing from the pure 1D system, we have chosen the values $r_0=1.2$ and $\epsilon=\epsilon_{\mathrm{max}}=0.8$, in which case $\lambda_{\clp}\simeq 1.67$. The open-source C++ code employed to obtain the results in this section is available for access through  Ref.~\cite{SingleFile3}.

\begin{figure}
	\includegraphics[width=\columnwidth]{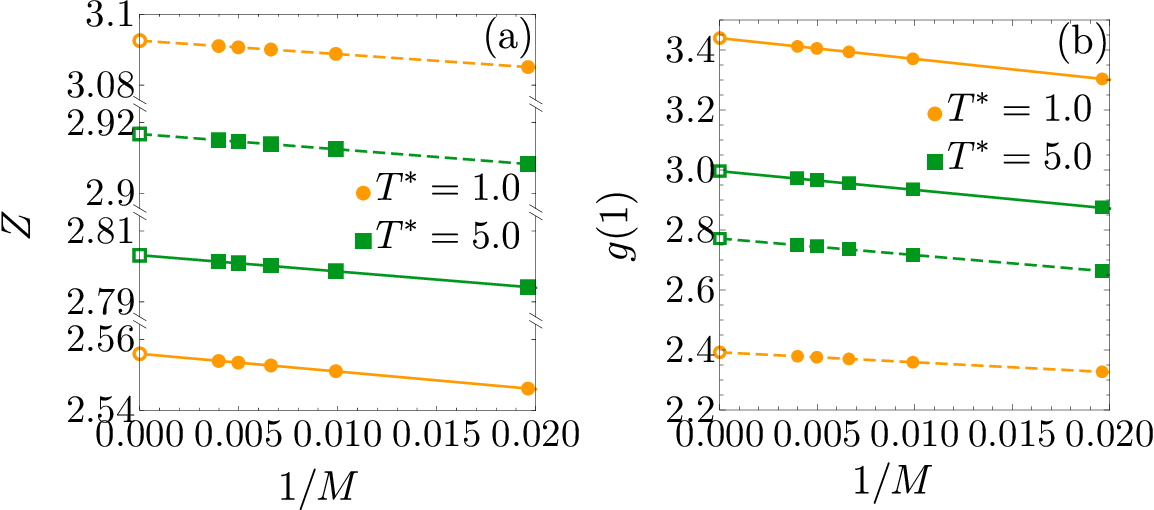}
	\caption{Plot of (a) $Z$ and (b) $g(1)$ versus $1/M$ for $T^*=1$ (circles) and $T^*=5$ (squares), in both cases with $\lambda=1$. The lines (solid for SW and dashed for SS) are linear fits to the numerical data. The open symbols at $1/M$ denote the extrapolations to $M\to\infty$.
}
	\label{fig:1/M}
\end{figure}

An  observation is worth mentioning. When delving into the theoretical expressions presented in Secs.~\ref{sec:properties} and \ref{sec:properties_g}, it becomes imperative to assign a finite value to $M$. As emphasized in Ref.~\cite{MS23}, opting for $M=251$ typically proves sufficiently large to render finite-$M$ effects practically negligible. Conversely, to eliminate any potential impact of a finite $M$, we systematically computed the relevant quantities for various $M$ values (specifically, $M=51,101,151,201,251$), modeled them as linear functions of $1/M$, and ultimately approached the limit $1/M\to 0$ in the extrapolations.
This procedure is illustrated in Fig.~\ref{fig:1/M} for  $Z$ and $g(1)$ at  $T^*=1$ and $5$ with $\lambda=1$. As observed, the local values of the RDF are notably more sensitive to finite $M$ than the thermodynamic quantities.

\subsection{Transverse density profile}

\begin{figure}
	\includegraphics[width=\columnwidth]{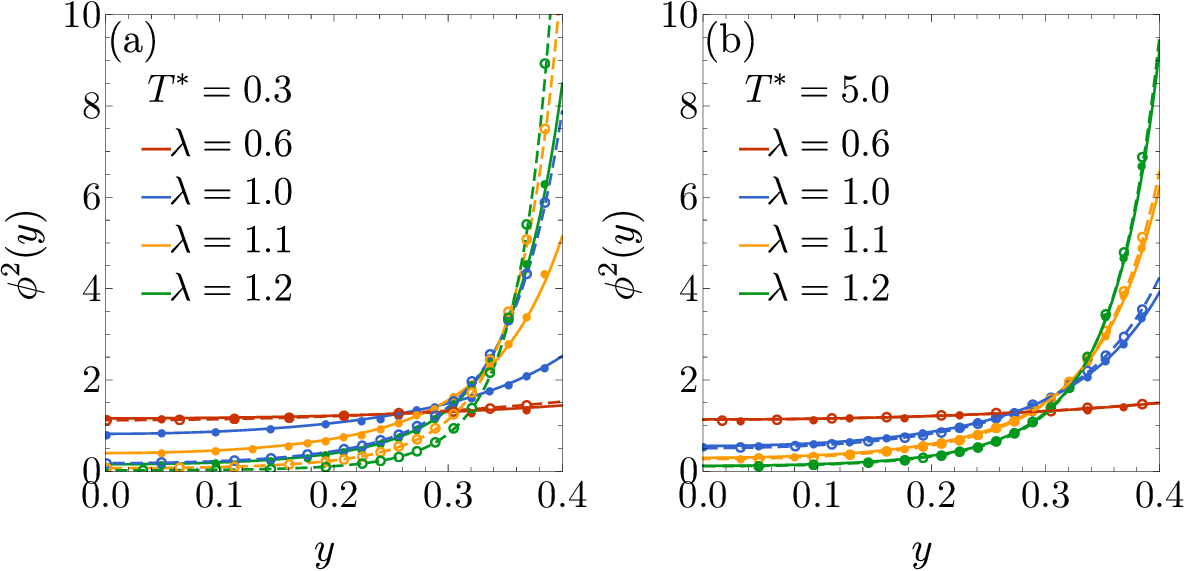}
	\caption{Transverse density profiles at  $\lambda=0.6$, $1.0$, $1.1$, and $1.2$  (from top to bottom in the region near $y=0$) for (a) $T^*=0.3$ and (b) $T^*=5.0$. Solid and dashed lines represent the SW and SS  systems, respectively. Symbols (closed for SW, open for SS) are MC simulation results.}
	\label{fig:fy}
\end{figure}

The transverse density profile $\phi^2(y)$, computed from Eq.~\eqref{eq:eigenproblem2}, is shown in Fig.~\ref{fig:fy} for both potentials at different densities and temperatures. In general, the density profile tends to be almost uniform at low densities, but becomes more abrupt, with more particles near the walls and fewer in the center of the pore, as the density increases. As close packing is approached, all particles tend to arrange in a zigzag configuration at both the top and bottom walls of the channel. Figure \ref{fig:fy}(a) shows that, at low temperatures and medium or high densities, the profiles are sharper in the case of the SS potential, where the excluded volume effects  are more dominant. At high temperatures, however, SW and SS fluids are nearly equivalent, as both behave essentially like HD fluids.

\subsection{Equation of state and excess internal energy}

\begin{figure}
	\includegraphics[width=\columnwidth]{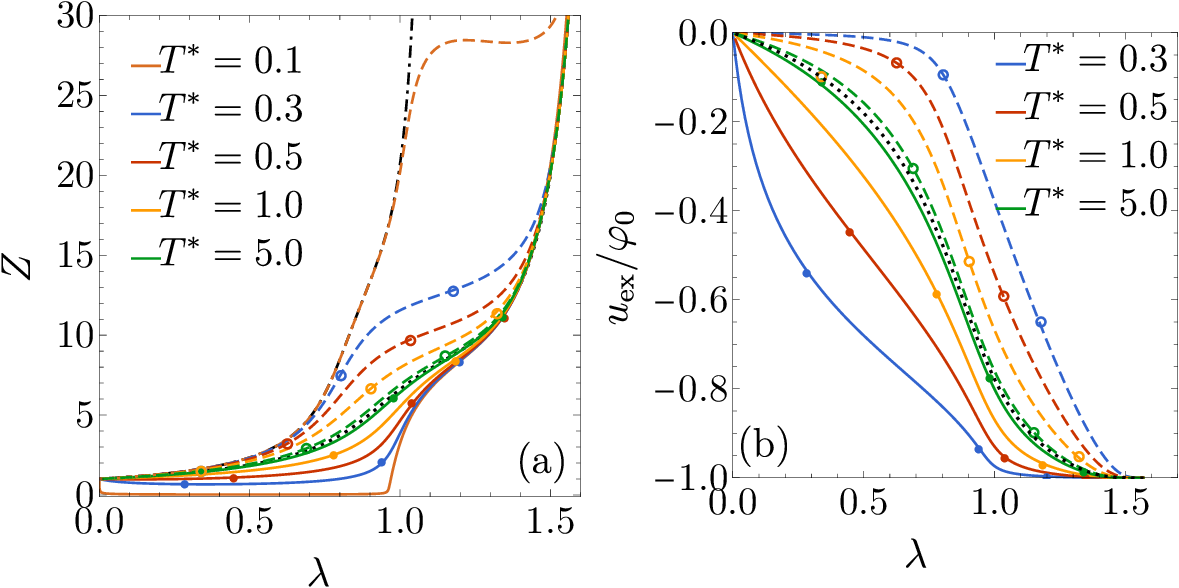}
	\caption{(a) Compressibility factor and (b) excess internal energy as functions of density at different representative temperatures for SW (solid lines) and SS (dashed lines) potentials. The black dotted lines represent the limit at infinite temperature (HD fluid with a hard-core diameter of $1$ and an excess pore width of $\epsilon=0.8$), while the black dash-dotted line in panel (a) represents the limit of the SS fluid at zero temperature (HD fluid with a diameter of $r_0=1.2$ and an excess pore width $\epsilon/r_0\simeq 0.67$). The temperatures are (from top to bottom in the SS case and from bottom to top  in the SW case) (a)  $T^*=0.1$, $0.3$, $0.5$, $1.0$, and $5.0$, and (b)  $T^*=0.3$, $0.5$, $1.0$, and $5.0$.
Symbols represent MC simulation results.}
	\label{fig:eos}
\end{figure}

The compressibility factor, Eq~\eqref{eq:eos22}, for different temperature values is shown in Fig.~\ref{fig:eos}(a). In the SW case, due to the attractive part of the potential, there exists a range of temperatures, $0<T^*<T_{\mathrm{B}}^*\simeq 0.59$ [see Eq.~\eqref{TBoyle}], for which $Z < 1$ at low densities, whereas  $Z>1$ is always fulfilled for every value of temperature and density in the SS case.

In agreement with Eq.~\eqref{T_infinity}, in the limit $T^* \rightarrow \infty$, both SW and SS fluids recover the equation of state of a confined HD fluid of unit diameter and pore width $\epsilon=0.8$, as can be observed in Fig.~\ref{fig:eos}(a).
As expected, at high densities and a nonzero temperature, the compressibility factor of both systems tends to that of a HD fluid, with $Z$ diverging at $\lambda=\lambda_{\clp}\simeq 1.67$.
It is also observed that, in agreement with Eq.~\eqref{T_zero}, in the SS case at zero temperature ($T^* \rightarrow 0$) we also recover the solution of a confined HD  system, where the disks have a hard-core diameter $r_0=1.2$, the excess pore width is $\epsilon/r_0\simeq 0.67$, and the density is $\lambda r_0=1.2\lambda$. Therefore, in the limit $T^* \rightarrow 0$, the compressibility factor of the SS fluid does not diverge at $\lambda_\clp=1/a(\epsilon)\simeq 1.67$, but at a smaller value $\lambda_\clp'=1/r_0a(\epsilon/r_0)=1/b(\epsilon)\simeq 1.12$.
If $T^*$ is small but nonzero, as is the case with $T^* = 0.1$ in Fig.~\ref{fig:eos}(a), the SS compressibility factor is practically indistinguishable from that at zero temperature for densities smaller than $\lambda_\clp' \simeq 1.12$. However, for higher densities, the curve deviates from the zero-temperature one and diverges at the true close-packing value $\lambda_\clp \simeq 1.67$.

The excess internal energy per particle, as derived from Eq.~\eqref{eq:uex22}, is shown in Fig.~\ref{fig:eos}(b) in units of $\varphi_0$ for both the SW potential, where $u_{\text{ex}}$ is always negative due to the attractive well ($\varphi_0>0$), and for the SS potential, where it is always positive ($\varphi_0<0$).
As density increases, $u_{\text{ex}}/\varphi_0$ tends to $-1$ since the coronas of neighbor particles are overlapped in the high-density regime. This effect is more pronounced for lower temperatures in the SW case. In contrast,  it is more accentuated for higher temperatures in the SS case because overpassing the repulsive barrier requires high enough temperatures. The solid dotted line in Fig.~\ref{fig:eos}(b) actually represents a \emph{nominal} excess energy for a HD fluid since it is obtained from Eq.~\eqref{eq:uex22} by using the HD eigenvalue $\ell$ and eigenfunction $\phi(y)$, even though $b(\Delta y)$ keeps being defined by Eq.~\eqref{b(s)}.

While not included in Fig.~\ref{fig:eos}, we have checked that, despite their simplicity, the approximations given by Eq.~\eqref{Z&u_approx} perform generally well for low to moderate densities. For instance, at $\lambda= 0.5$, the relative deviations in the SW (SS) compressibility factor are $99\%$ ($0.6\%$), $57\%$ ($1\%$), $27\%$ ($1\%$), $8\%$ ($2\%$), and $0.2\%$ ($1\%$) for $T^*=0.1$, $0.3$, $0.5$, $1$, and $5$, respectively. The respective deviations in the excess internal energy are
$1\%$ ($12\%$), $29\%$ ($11\%$), $33\%$ ($8\%$), $25\%$ ($1\%$), and $13\%$ ($8\%$).
Note that the large deviations  in $Z$ for the low-temperature SW fluid are due to the small values of $Z$ at $\lambda=0.5$, specifically $Z=0.030$ and $Z=0.69$ for $T^*=0.1$ and $0.3$, respectively.

\begin{figure}
	\includegraphics[width=\columnwidth]{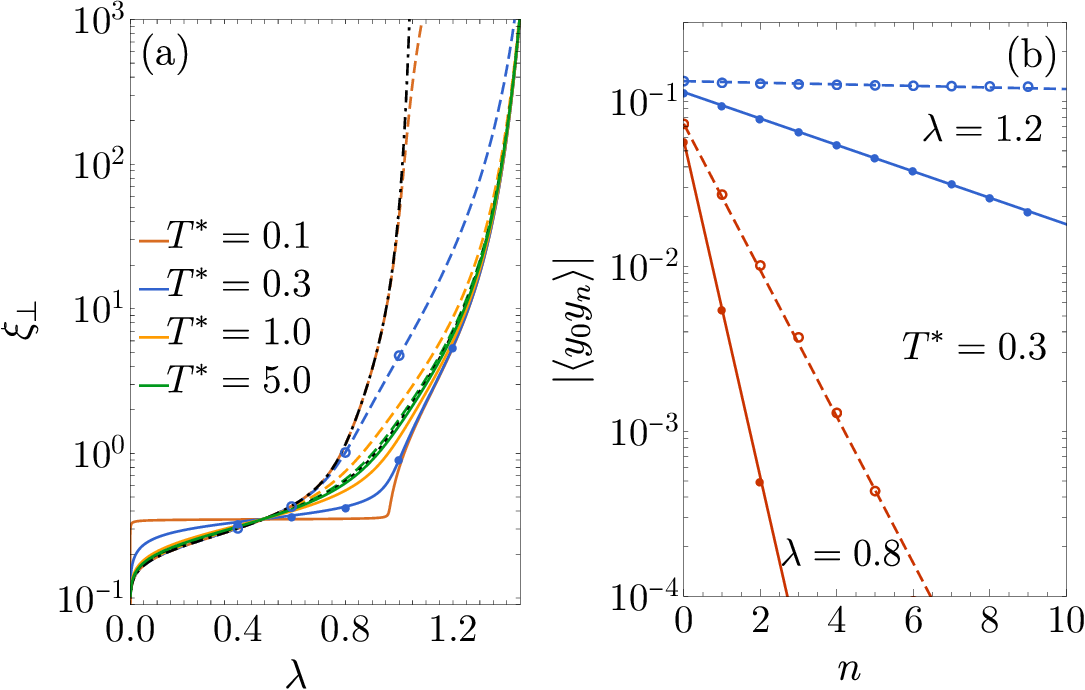}
	\caption{(a) Transverse correlation degree as a function of density at different representative temperatures for SW (solid lines) and SS (dashed lines) potentials.
The black dotted line represents the limit at infinite temperature (HD fluid with a hard-core diameter of $1$ and an excess pore width of $\epsilon=0.8$), while the black dash-dotted line represents the limit of the SS fluid at zero temperature (HD fluid with a diameter of $r_0=1.2$ and an excess pore width $\epsilon/r_0\simeq 0.67$).
The temperatures are (from top to bottom in the SS case and from bottom to top  in the SW case)  $T^*=0.1$, $0.3$, $1.0$, and $5.0$.
Symbols represent MC simulation results. (b) Illustration of the evaluation of $\xi_\perp$  in simulations from the slope of $|\langle y_0y_n\rangle|$ (in logarithmic scale) vs $n$.}
	\label{fig:xi_perp}
\end{figure}

\subsection{Transverse correlation degree}
The transverse correlation degree $\xi_\perp$ is plotted in Fig.~\ref{fig:xi_perp}(a) as a function of the linear density at $T^*=0.1$, $0.3$, $1.0$, and $5.0$. Figure~\ref{fig:xi_perp}(b) illustrates the behavior of $|\langle y_0 y_n\rangle|$ and the evaluation of $\xi_\perp$ in our MC simulations.

At a given density, $\xi_\perp$ increases with increasing temperature in the SW case, while the opposite trend is present in the SS case. In the limit $T^*\to\infty$ both the SW and SS curves collapse to the curve corresponding to the HD interaction, while a related collapse occurs in the limit $T^*\to 0$ for the SS fluid.
We observe that $\xi_\perp<0.3$ if $\lambda<0.5$. This indicates that the transverse coordinates of first-neighbor particles are minimally correlated within this regime. However, $\xi_\perp$ rapidly increases with increasing density, indicating that the transverse positions of distant neighbors become progressively more correlated.

\begin{figure}
	\includegraphics[width=\columnwidth]{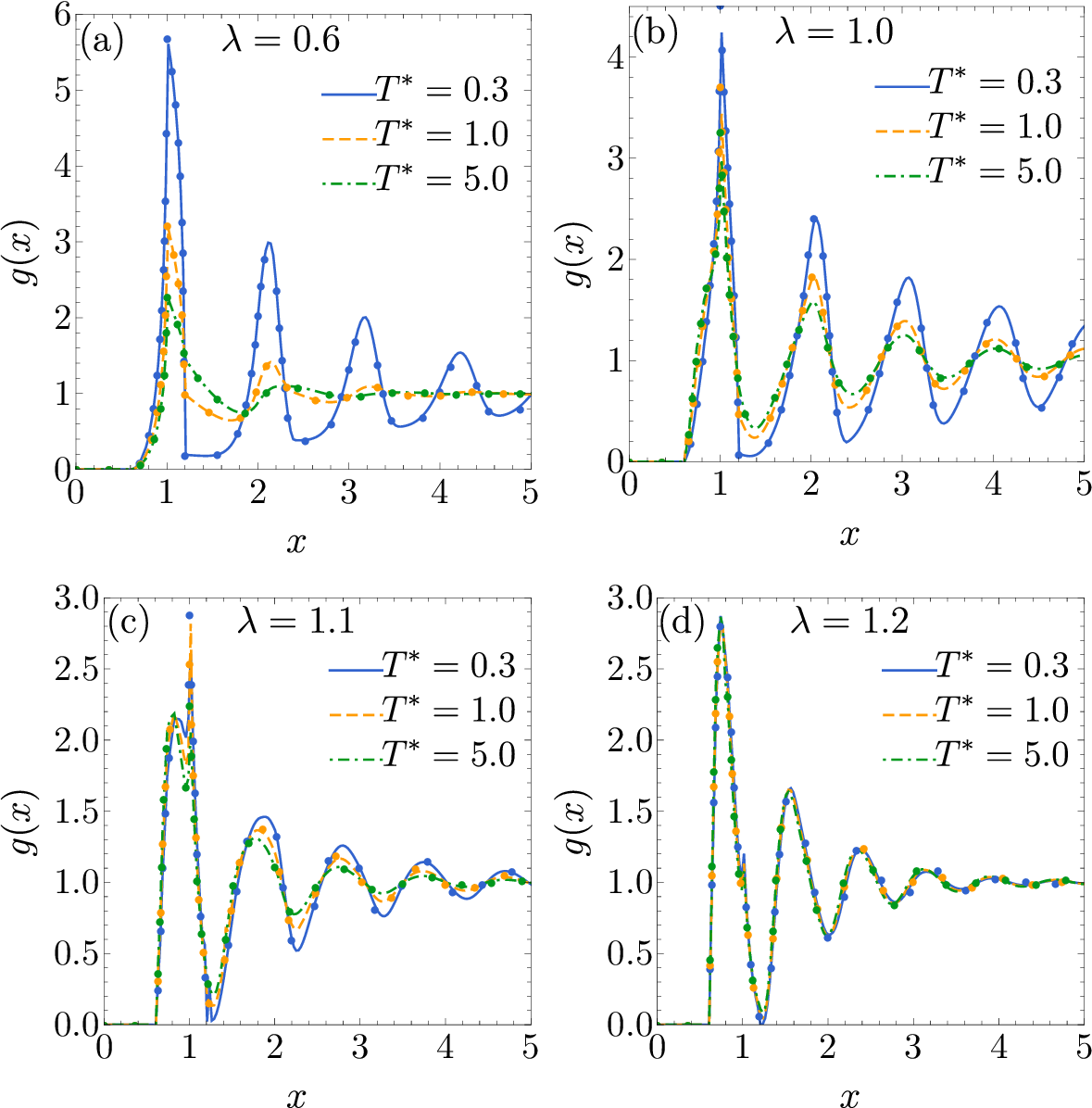}
	\caption{Total RDF for the SW fluid at different temperatures for several values of density:  (a) $\lambda=0.6$, (b) $\lambda=1.0$, (c) $\lambda=1.1$, and (d) $\lambda=1.2$. Symbols are MC simulation results.}
	\label{fig:gxSW}
\end{figure}

\begin{figure}
	\includegraphics[width=\columnwidth]{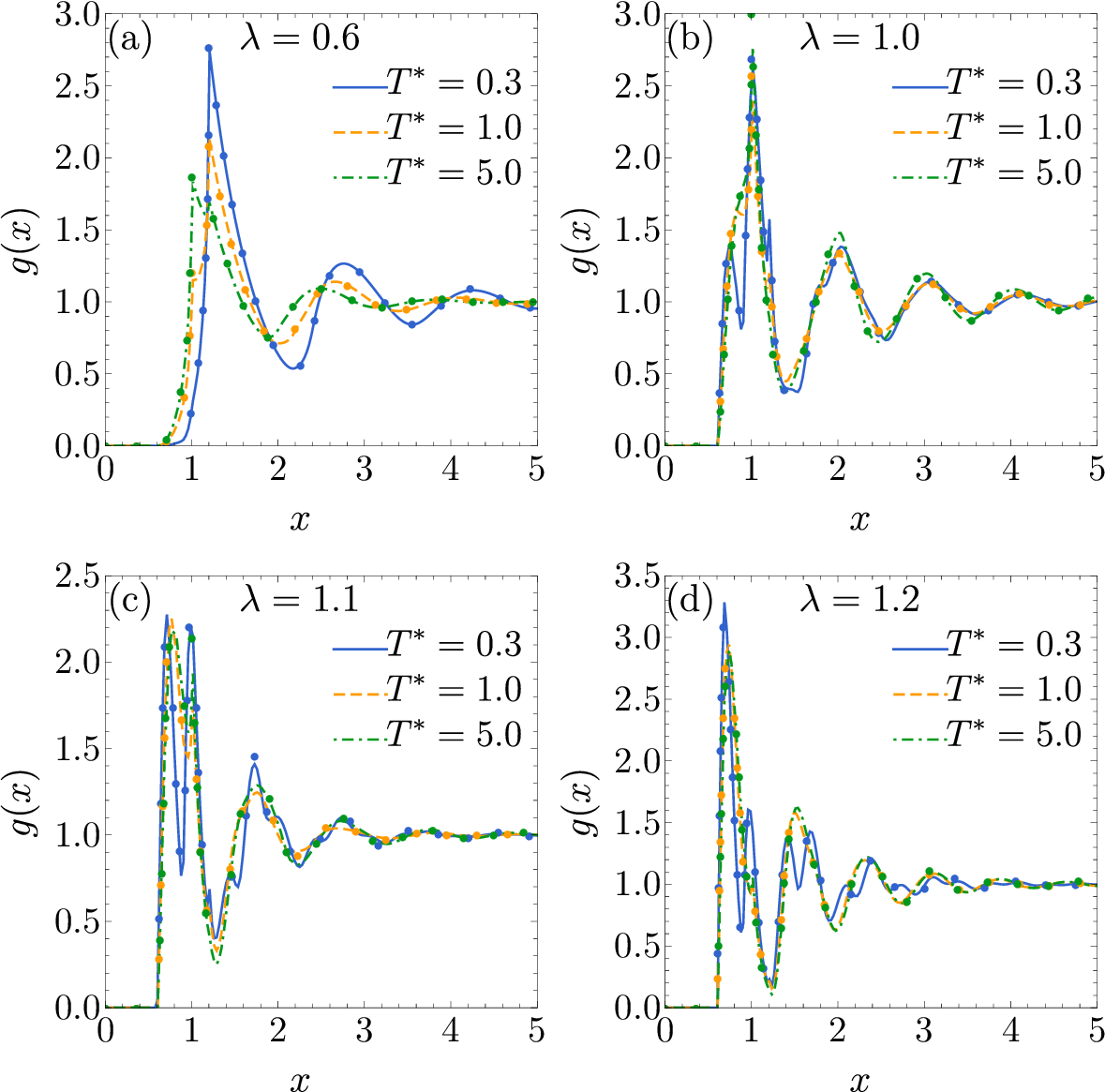}
	\caption{Total RDF for the SS fluid at different temperatures for several values of density:  (a) $\lambda=0.6$, (b) $\lambda=1.0$, (c) $\lambda=1.1$, and (d) $\lambda=1.2$. Symbols are MC simulation results.}
	\label{fig:gxSS}
\end{figure}

\subsection{Radial distribution function}

\subsubsection{Total RDF}

The RDF is one of the most important structural quantities in any system, as it measures how the local density around a reference particle varies with distance.

In Fig.~\ref{fig:gxSW}, the total RDF for the SW potential is illustrated across varying densities and temperatures. Notably, at lower densities, temperature emerges as a key factor influencing the amplitude of the oscillations. However, this dependency diminishes substantially at higher densities, where the RDF undergoes minimal alteration with temperature variations, resembling closely the RDF of the HD fluid at equivalent density.
The positions of the minima and maxima are particularly influenced by density but exhibit minimal sensitivity to temperature changes. Specifically, our observations indicate that the first peak occurs around $x=1$ at $\lambda=0.6$ and $\lambda=1.0$, while a local maximum emerges near $x=a(\epsilon)=0.6$ at $\lambda=1.1$. Notably, this local maximum becomes the absolute maximum at $\lambda=1.2$.

For the SS potential, the RDF is presented in Fig.~\ref{fig:gxSS}, with the same densities and temperatures as depicted in Fig.~\ref{fig:gxSW}. Due to the repulsive nature of the potential, temperature plays a larger role in the position of the peaks than in the SW case, especially at low densities ($\lambda=0.6$), where the first peak shifts from $x=r_0=1.2$ to $x=1.0$ with increasing temperature. At higher densities, lower temperatures result in a significantly less ordered structure.
For $\lambda=1.0$ and $T^*=5.0$, the peak of $g(x)$ is located at $x \approx 1$. However, at lower temperatures ($T^*=1.0$ and $T^*=0.3$), a secondary peak appears near $x = a(\epsilon) \approx 0.6$. When the density is increased to $\lambda=1.1$, the peak at $x \approx a(\epsilon)$ becomes more prominent, while the peak at $x \approx 1$ becomes secondary and then disappears at $\lambda=1.2$, except if $T^*=0.3$.
This phenomenology is consistent with the observation that, as density increases, the structural properties of the SW and SS fluids progressively resemble those of the HD fluid. This tendency is more pronounced at higher temperatures.

\subsubsection{Partial RDFs}

In contrast to the total RDF, partial RDFs describe spatial correlations between particles at \emph{fixed} transverse coordinates. Out of all possible partial RDFs, $g(y,y';x)$, the most interesting ones correspond to $y,y'=\pm\ehalf$, that is,
\begin{subequations}
\begin{equation}
	g_{++}(x) \equiv g\left(\ehalf,\ehalf;x\right) = g\left(-\ehalf,-\ehalf;x\right),
\end{equation}
\begin{equation}
	g_{+-}(x) \equiv g\left(\ehalf,-\ehalf;x\right) = g\left(-\ehalf,\ehalf;x\right).
\end{equation}
\end{subequations}
While $g_{++}(x)$ encodes spatial correlations between two particles both located at the top (or bottom) part of the channel, $g_{+-}(x)$ measures the spatial correlations between a particle in contact with one wall and a particle in contact with the opposite wall. Note that near close packing, all particles are very close to the walls, so that $g(x) \simeq \frac{1}{2}[g_{++}(x) + g_{+-}(x)]$.

\begin{figure}
	\includegraphics[width=\columnwidth]{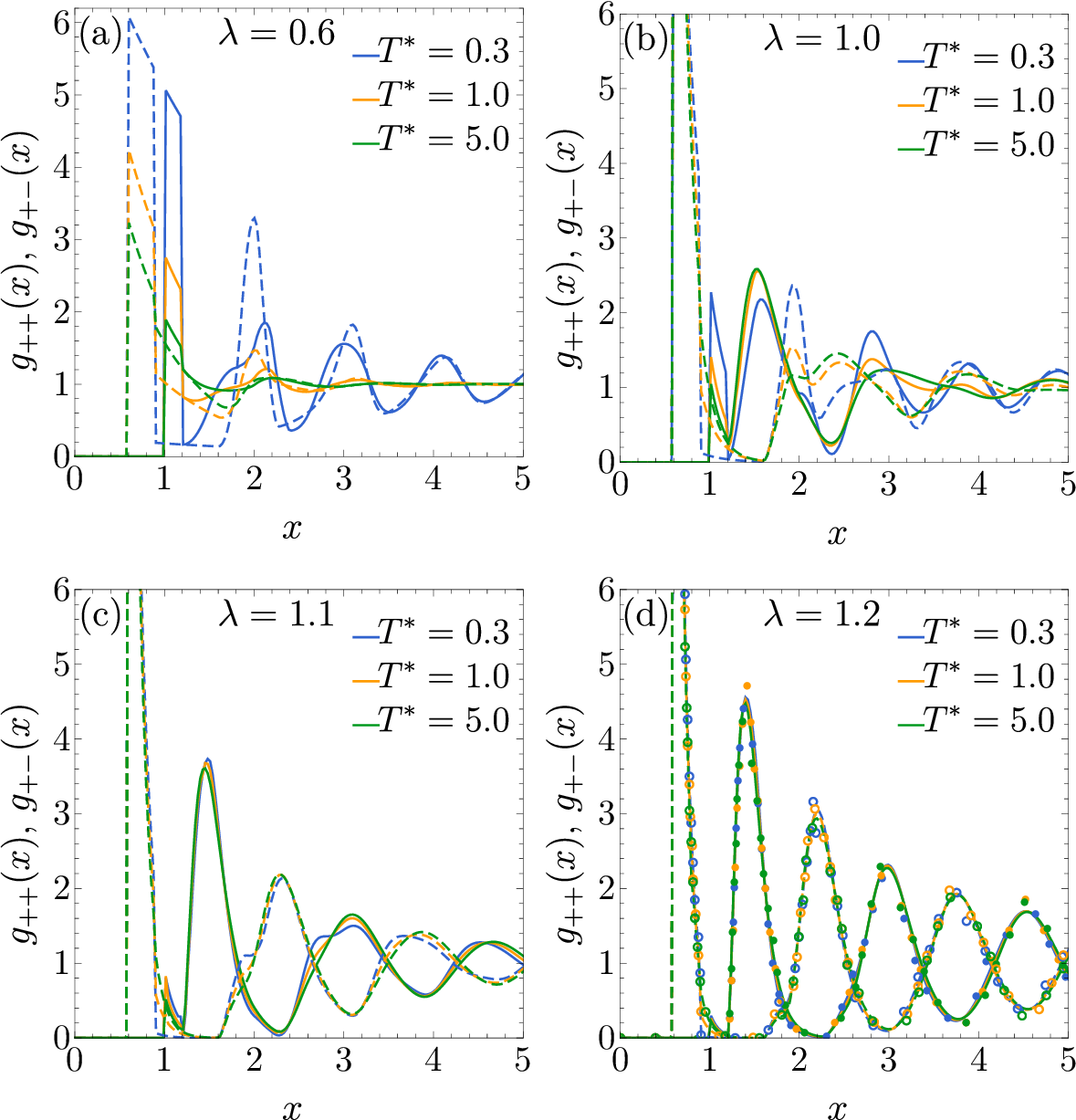}
	\caption{Partial RDFs  $g_{++}(x)$ (solid lines) and $g_{+-}(x)$ (dashed lines) for the SW fluid at different temperatures for several values of density:  (a) $\lambda=0.6$, (b) $\lambda=1.0$, (c) $\lambda=1.1$, and (d) $\lambda=1.2$. Symbols in panel (d) are MC simulation results. Note that the oscillations tend to  become more pronounced as $T^*$ decreases.}
	\label{fig:gxpartialSW}
\end{figure}

\begin{figure}
	\includegraphics[width=\columnwidth]{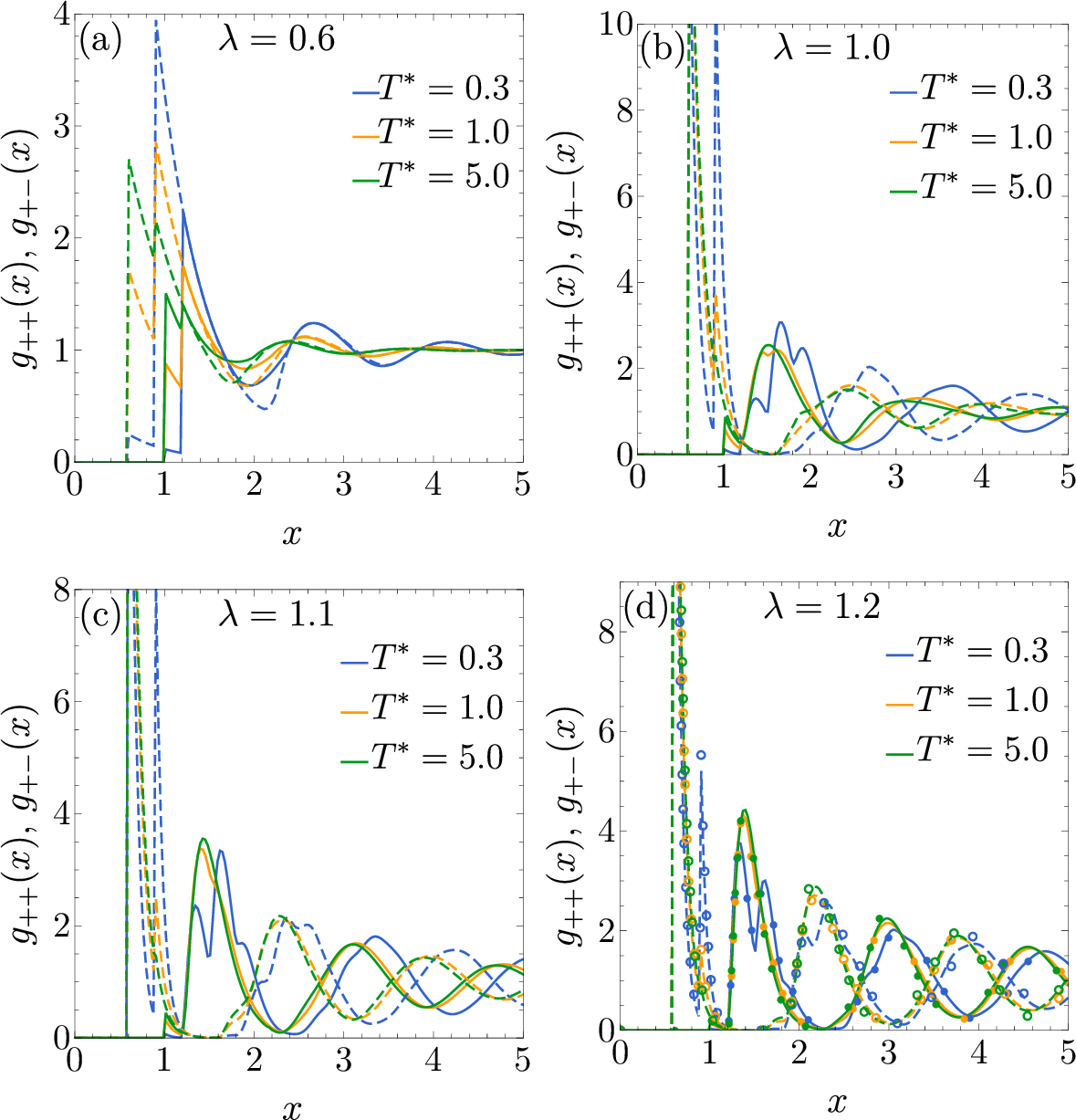}
	\caption{Partial RDFs  $g_{++}(x)$ (solid lines) and $g_{+-}(x)$ (dashed lines) for the SS fluid at different temperatures for several values of density:  (a) $\lambda=0.6$, (b) $\lambda=1.0$, (c) $\lambda=1.1$, and (d) $\lambda=1.2$. Symbols in panel (d)  are MC simulation results. Note that the oscillations tend to  become more pronounced as $T^*$ decreases.}
	\label{fig:gxpartialSS}
\end{figure}

Figures~\ref{fig:gxpartialSW} and~\ref{fig:gxpartialSS} show $g_{++}(x)$ and $g_{+-}(x)$ for the SW and SS potentials, respectively, at the same temperatures and densities as in Figs.~\ref{fig:gxSW} and \ref{fig:gxSS}.
We have included MC simulation data for the density $\lambda=1.2$ only because, for $\lambda \leq 1.1$, the accumulation of particles at the walls is not high enough (see Fig.~\ref{fig:fy}) to ensure good statistics in the evaluation of $g_{++}(x)$ and $g_{+-}(x)$.
In both classes of potentials, $g_{++}(x)= 0$ if $x<1$ and $g_{+-}(x) = 0$ if $x<a(\epsilon)=0.6$, as expected. Also in both cases, the disappearance of the peak in $g_{++}(1^+)$ when density is increased is  directly related to the disappearance of defects in the zigzag structure that arises in the close-packing configuration \cite{MS23b}.
In fact, we have checked that at $\lambda=1.5 = 0.90\lambda_\clp$ (not shown in Figs.~\ref{fig:gxpartialSW} and \ref{fig:gxpartialSS}), the functions $g_{++}(x)$ and $g_{+-}(x)$ are hardly distinguishable from those of a HD fluid, as displayed in Fig.~6 of Ref.~\cite{MS23b}.

\begin{figure}
	\includegraphics[width=\columnwidth]{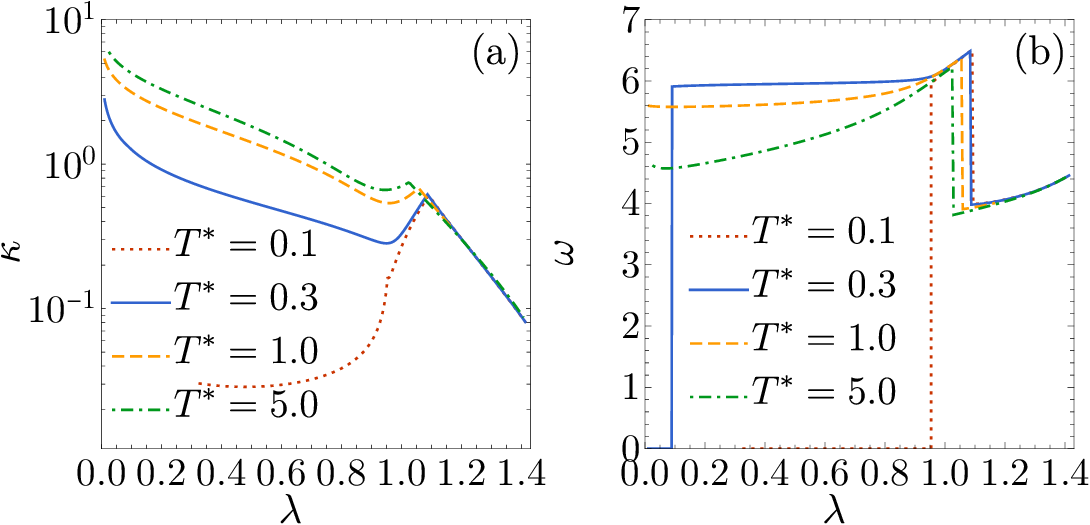}
	\caption{Plot of (a) the inverse correlation length and (b) the oscillation frequency as functions of density at different temperatures  for the SW fluid.}
	\label{fig:polesSW}
\end{figure}

\begin{figure}
	\includegraphics[width=\columnwidth]{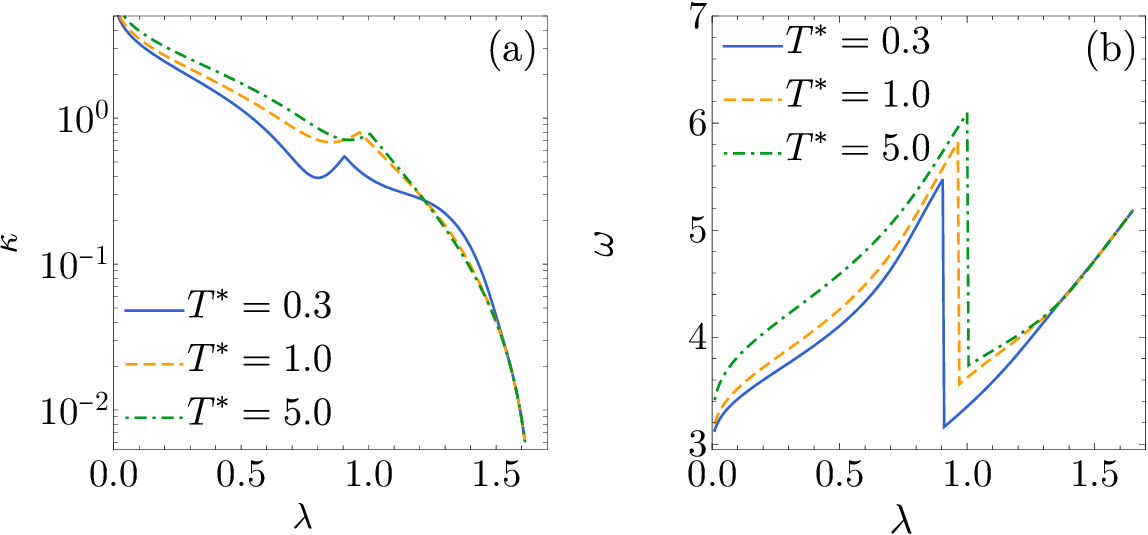}
	\caption{Plot of (a) the inverse correlation length and (b) the oscillation frequency as functions of density  at different temperatures for the SS fluid.}
	\label{fig:polesSS}
\end{figure}

\subsubsection{Asymptotic behavior}

As elaborated in Sec.~\ref{sec3E}, the large-$x$ asymptotic behavior of the RDF is determined by the dominant poles of $\widetilde{G}(y,y';s)$. To obtain them, we have started from the discrete version with finite $M$ [see Eq.~\eqref{eq:rdf}] and found the zeros of $\det\left[\mathsf{I}-A^2\mathsf{\Omega}(s+\beta p)\right]$  closest to the imaginary axis. Then, the limit $M\to\infty$ was taken by following the extrapolation method illustrated in Fig.~\ref{fig:1/M}.

Figures~\ref{fig:polesSW} and~\ref{fig:polesSS} show the evolution of the values of $\kappa$ and $\omega$ associated with the leading pole as  functions of density. The inverse correlation length, $\kappa$, is always continuous but the oscillation frequency, $\omega$, does present discontinuous jumps that correspond to structural changes. In the case of the SW potential [see Fig.~\ref{fig:polesSW}(b)], as density increases for very low temperatures ($T^*=0.1$ and $0.3$), a first discontinuous jump from $\omega=0$ to $\omega\neq 0$ represents a Fisher--Widom transition~\cite{FW69} from monotonic to oscillatory decay of $h(y,y';x)$ [see Eqs.~\eqref{eq:polereal} and \eqref{eq:poleimag}]. Although not apparent on the scale of Fig.~\ref{fig:polesSW}(b), this transition persists at very low densities for higher temperatures (e.g., $T^*=1$ and $T^*=5$).
The Fisher--Widom transition from monotonic to damped oscillatory decay signals a competition between the attractive and repulsive parts of the interaction \cite{SHRE19,MRYSH24}. Consequently, this transition is absent  in the SS fluid, regardless of temperature. However, a jump from a higher frequency $\omega_\text{I}$ to a smaller nonzero frequency $\omega_\text{II}$ takes place at $\lambda\approx 1$ for any temperature and both types of interaction.
The latter transition reflects a competition between the two distance scales ($1$ and $r_0$) in the interaction potential, as described in Eq.~\eqref{varphi(r)}.

The abrupt shifts in $\omega$ stem from the crossing of two competing poles with identical real parts, leading to distinctive kinks in $\kappa$.
At the density $\lambda \approx 1$ where the transition $\omega_\text{I} \leftrightarrow \omega_\text{II}$ occurs and $\kappa$ exhibits a kink, the asymptotic behavior of $h(y,y';x)$ is of the form $\sim e^{-\kappa x}\left[\cos(\omega_\text{I} x + \delta_\text{I}) + \mathcal{C}\cos(\omega_\text{II} x + \delta_\text{II})\right]$, where $\mathcal{C}$ is the ratio between the two amplitudes. Analogously, at the Fisher--Widom transition $\omega=0 \leftrightarrow \omega\neq 0$ in the SW case, one has
$h(y,y';x)\sim e^{-\kappa x}\left[\cos(\omega x + \delta) + \mathcal{C}\right]$. However, these transitions and kinks of $\kappa$ do not manifest  in the thermodynamic quantities.

\begin{figure}
	\includegraphics[width=0.7\columnwidth]{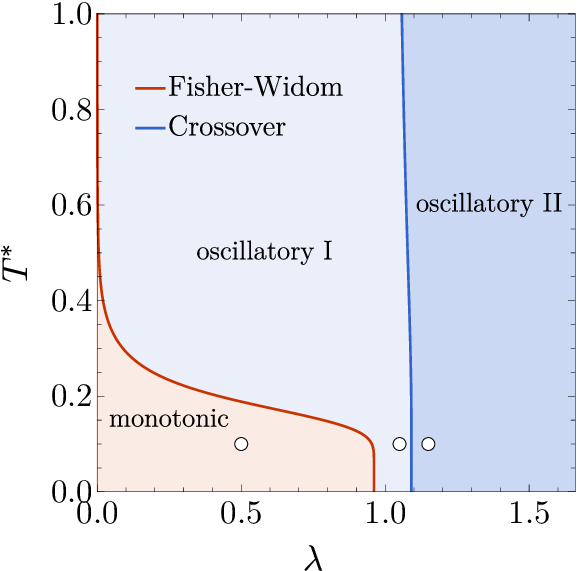}
	\caption{Phase diagram for the SW fluid on the plane $T^*$ vs $\lambda$. The circles represent the states considered in Fig.~\ref{fig:grPoles}.}
	\label{fig:linesSW}
\end{figure}

The phase diagram illustrating the types of asymptotic decay of $h(y,y';x)$ for the SW fluid is presented in Fig.~\ref{fig:linesSW}. Three distinct regions can be discerned on the $T^*$ vs $\lambda$ plane. For densities less than $\lambda\simeq 0.9612$ and sufficiently low temperatures, the decay is exclusively monotonic, owing to the prevailing influence of the attractive part of the interaction potential. This region is demarcated from the oscillatory decay region by the Fisher--Widom line~\cite{FW69}. Subsequently, the oscillatory decay region is partitioned into two subregions by a crossover line~\cite{SPTER16}. Upon traversing this crossover line with increasing density, the oscillation frequency undergoes a sudden transition from a value $\omega_{\text{I}}\approx 2\pi\simeq 6.3$ (oscillatory region I) to a smaller value $\omega_{\text{II}}\approx 4$ (oscillatory region II), mirroring the behavior observed in the HD case~\cite{MS23b}.

The transition between oscillatory regions I and II occurring at $\lambda \approx 1$ may be linked to recent discussions about the critical role of this value \cite{PBT23}. At a given excess pore width $\epsilon > 0$, configurations can be strictly linear if $\lambda < 1$, whereas configurations must exhibit a certain zigzag ordering if $\lambda > 1$. This might explain the sudden change in oscillation frequency at $\lambda \approx 1$.

\begin{figure}
	\includegraphics[width=\columnwidth]{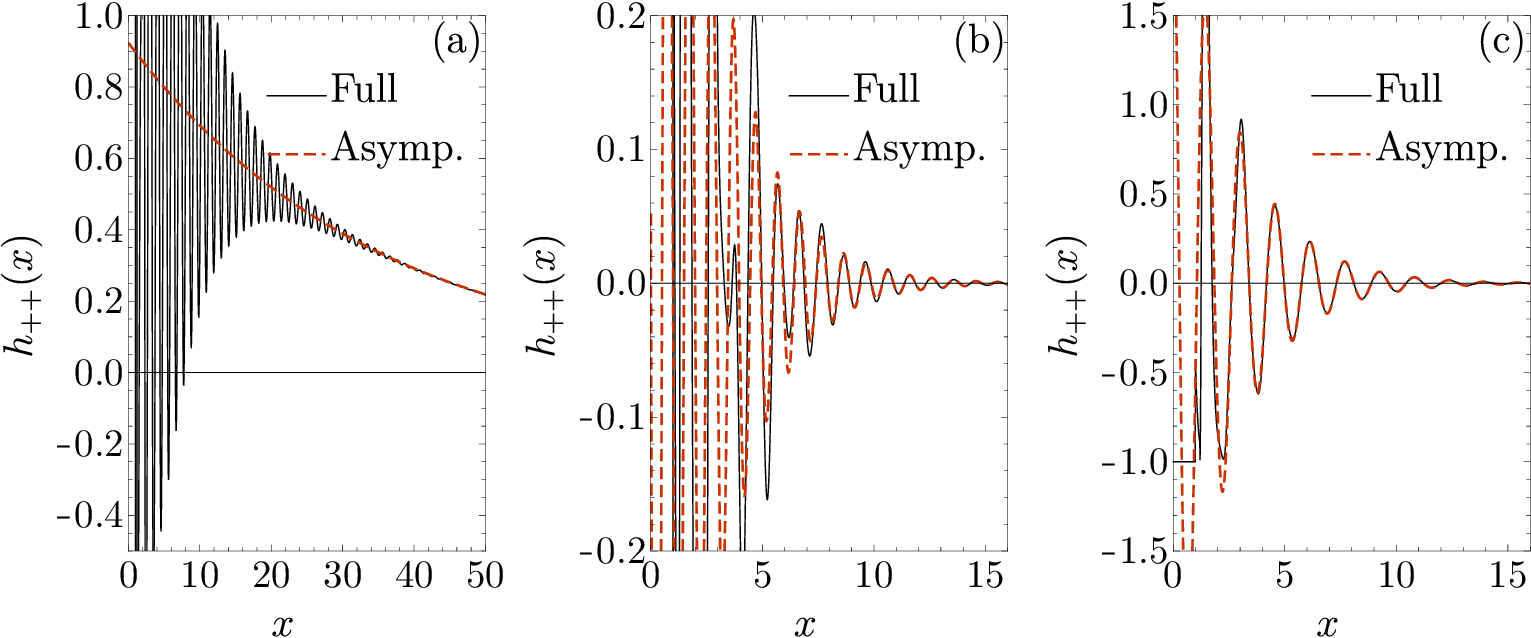}
	\caption{Plot of $h_{++}(x)$ for the SW fluid with $T^*=0.1$ and  (a) $\lambda=0.5$, (b) $\lambda=1.05$, and (c) $\lambda=1.15$ (see circles in Fig.~\ref{fig:linesSW}). The solid lines correspond to the full functions, while the dashed lines represent the asymptotic behaviors [see Eqs.~\eqref{eq:poleimag} and \eqref{eq:polereal}].}
	\label{fig:grPoles}
\end{figure}

\begin{figure}
	\includegraphics[width=0.7\columnwidth]{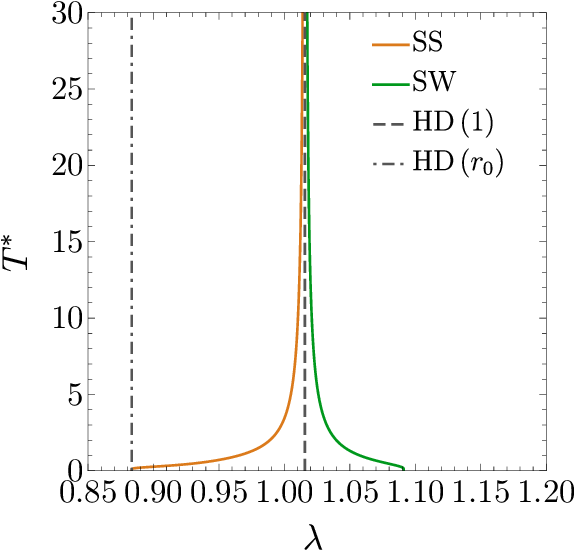}
	\caption{Structural crossover lines delineating transitions between two distinct oscillation frequencies (oscillatory regions I and II) are depicted for the SW fluid (right curve) and SS fluid (left curve). The dashed and dash-dotted vertical lines indicate the crossover densities at $\lambda=1.016$ and $\lambda=1.060/r_0=0.883$, respectively, corresponding to two confined HD fluids: HD~(1), characterized by a hard-core diameter of $1$ and an excess pore width of $\epsilon=0.8$, and HD~($r_0$), with a diameter of $r_0=1.2$ and an excess pore width $\epsilon/r_0\simeq 0.67$.}
	\label{fig:CRLines}
\end{figure}

To corroborate the insights obtained from the leading-pole analysis as applied to the SW case, Fig.~\ref{fig:grPoles} juxtaposes the complete total correlation function $h_{++}(x)$ with its asymptotic expressions derived from Eqs.~\eqref{eq:poleimag} and \eqref{eq:polereal}. The comparison is conducted for three particular states identified with circles in Fig.~\ref{fig:linesSW}, specifically $T^*=0.1$ and $\lambda=0.5$, $1.05$, and $1.15$.
The convergence of the complete functions to the anticipated asymptotic forms for extended distances is evident. In instances of asymptotic monotonic decay, as illustrated in Fig.~\ref{fig:grPoles}(a), the agreement necessitates a more extended range of distances compared to cases where the decay exhibits oscillations, whether with a higher frequency [Fig.~\ref{fig:grPoles}(b)] or a lower frequency [Fig.~\ref{fig:grPoles}(c)].

As mentioned earlier, the purely repulsive SS system lacks a FW line but exhibits crossover transitions between two distinct oscillation frequencies (see Fig.~\ref{fig:polesSS}). The crossover lines for the SW and SS fluids are presented in Fig.~\ref{fig:CRLines}. With increasing temperature, both lines converge toward the crossover density ($\lambda=1.016$) of the HD fluid with a unit diameter and the same excess pore width $\epsilon=0.8$, consistent with the general property indicated by Eq.~\eqref{T_infinity}. Additionally, at the opposite low-temperature limit, the SS line terminates at $\lambda=1.060/r_0=0.883$, aligning with the expected value for a HD system comprising particles with a diameter of $r_0=1.2$ and an excess pore width $\epsilon/r_0\simeq 0.67$, as predicted by Eq.~\eqref{T_zero}.

We have observed that, in the high-density regime, the asymptotic oscillations of $h_{++}(x)$ and $h_{+-}(x)$ are out of phase by half a wavelength. As a consequence the asymptotic behavior of $h(x)$ is governed by the subdominant pole.

\subsection{Structure factor}

\begin{figure}
	\includegraphics[width=0.95\columnwidth]{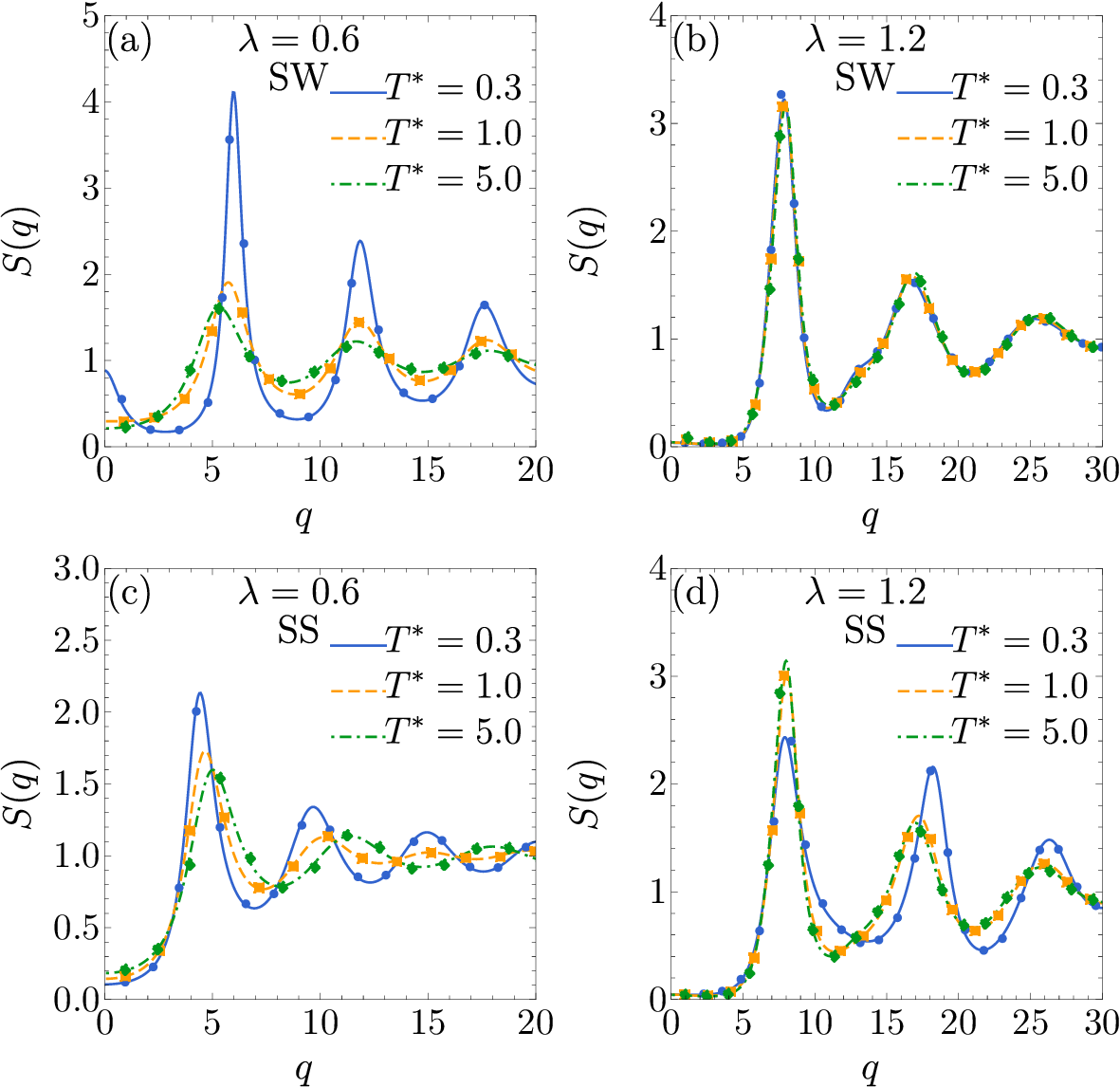}
	\caption{Structure factor at different temperatures and for densities (a, c) $\lambda=0.6$ and (b, d) $\lambda=1.2$. Panels (a, b) pertain to the SW fluid, while panels (c, d) pertain to the SS fluid. Symbols are MC simulation results.}
	\label{fig:SqAll}
\end{figure}

The importance of the structure factor lies in the fact that it is directly related to the intensity of radiation scattered by the fluid and can be therefore directly accessed via scattering experiments. Figure~\ref{fig:SqAll} shows the structure factor for several representative densities and temperatures for the SW and SS systems. In general, relative maxima are closer to one another at low densities, while they become more spaced out with increasing density. In parallel with what was observed in Figs.~\ref{fig:gxSW} and \ref{fig:gxSS}, the role of temperature is more important at low densities than at high densities, especially in the case of the SW potential. In the latter case, the structure factor at high density is practically independent of temperature.

\section{\label{sec:concluding}Concluding remarks}

In this study, we have investigated the impact of attractive and repulsive coronas on hard-core disks within confined geometries. Employing the SW and SS pairwise interactions between disks confined in an extremely narrow channel (q1D configuration), we have precisely examined their thermodynamic and structural properties. This exploration is facilitated through an exact mapping of the q1D system onto a nonadditive polydisperse mixture of rods with equal chemical potential, allowing for a detailed analysis of the system's behavior.

Our initial focus was on investigating the fundamental thermodynamic properties, including the equation of state and excess internal energy. We explored their dependence on density and temperature while examining their limiting behaviors at both very high and low temperatures. Additionally, we derived the second virial coefficient and determined the Boyle temperature for the SW potential, providing a comprehensive understanding of the system's thermodynamic characteristics.

Furthermore, we delved into the structural properties, encompassing the RDF, both total and partial, and the structure factor. An analytical expression for the RDF at short distances was successfully derived. Our investigation extended to the asymptotic large-distance behavior, where we computed the correlation length and the oscillation frequency of the RDF. The results demonstrated a full consistency with the complete functions, underscoring the robustness of our analytical approach in capturing the system's structural characteristics across various length scales.

While phase transitions do not manifest in these q1D systems, our investigation revealed discontinuous structural changes concerning the asymptotic oscillation frequency for both potentials. Additionally, the FW line, characterizing the transition from monotonic to oscillatory asymptotic decay in the SW system, was identified. These findings highlight subtle yet significant structural transformations in the system's behavior, enriching our understanding of its complex dynamics in confined geometries.

To affirm the accuracy of the q1D$\rightarrow$1D mapping, we conducted NPT and NVT MC simulations of the actual 2D system. The comparison between the theoretical predictions and the simulation results serves as a robust confirmation of the fidelity of our mapping approach, enhancing the reliability of our theoretical predictions in capturing the features of the true confined 2D system.

While the emphasis of this paper has been on longitudinal properties, it is noteworthy that the mapping technique employed enables the derivation of transverse properties as well. A detailed exploration of these transverse properties is presented in a separate work \cite{MS24b}, providing a comprehensive examination of the system's behavior in both longitudinal and transverse dimensions.

\acknowledgments
We express our gratitude to R.~Fantoni for generously sharing preliminary versions of Ref.~\cite{F23} and for engaging in insightful discussions about the project that inspired this paper.
Financial support from Grant No.~PID2020-112936GB-I00 funded by the Spanish agency MCIN/AEI/10.13039/501100011033 and from Grant No.~IB20079 funded by Junta de Extremadura (Spain) and by ERDF
``A way of  making Europe'' is acknowledged.
A.M.M. is grateful to the Spanish Ministerio de Ciencia e Innovaci\'on for support from a predoctoral fellowship Grant No.~PRE2021-097702.

\appendix

\section{On the mapping q1D$\leftrightarrow$1D}
\label{app:q1D<->1D}
Let us consider a q1D system of 2D interacting particles subject to a an external wall potential that constraints them to single-file formations, such that any particle $\alpha$ can only interact with its two adjacent neighbors $\alpha-1$ and $\alpha+1$.
The longitudinal and transverse lengths of the system are $L$ and $\epsilon$, respectively.
For convenience, we consider here the grand canonical ensemble, whose associated partition function is \cite{S16}
\beq
\label{Xiq1D}
\Xi^{\text{q1D}}(\beta,L,\epsilon,\mu)=1+\sum_{N=1}^\infty \frac{ e^{\beta\mu N}}{\Lambda_{\mathrm{dB}}^{2N}} \mathcal{Q}_N^{\text{q1D}}(\beta,L,\epsilon),
\eeq
where the canonical configuration integral is
\bal
\label{Q_N}
\mathcal{Q}_N^{\text{q1D}}(\beta,L,\epsilon)=&\int_\epsilon dy_1 \int_\epsilon dy_2\cdots \int_\epsilon dy_N \int_0^L dx_1\int_{x_1}^L dx_2\nn
&\times  \cdots \int_{x_{N-1}}^L dx_N\, e^{-\beta\Phi_N(\{x_\alpha,y_\alpha\})}
\eal
and the total potential energy is
\beq
\label{Phi_N}
\Phi_N(\{x_\alpha,y_\alpha\})=\sum_{\alpha=1}^N \varphi(r_{\alpha,\alpha+1}),
\eeq
with $r_{\alpha,\alpha'}\equiv\sqrt{(x_\alpha-x_{\alpha'})^2+(y_\alpha-y_{\alpha'})^2}$ and where $\varphi(r)$ is the pair interaction potential. Note that, in Eq.~\eqref{Phi_N}, we have applied periodic boundary conditions in the longitudinal direction, so that $x_{N+1}=x_1+L$ and $y_{N+1}=y_1$.

To make the contact with a 1D system more direct, let us discretize the transverse coordinate as in Eq.~\eqref{y_i}. In that case, Eq.~\eqref{Q_N} becomes
\bal
\label{Q_N_2}
\mathcal{Q}_N^{\text{q1D}}(\beta,L,\epsilon)=&(\delta y)^N \sum_{i_1=1}^M\sum_{i_2=1}^M\cdots \sum_{i_N=1}^M\int_0^L dx_1\int_{x_1}^L dx_2\nn
&\times
 \ldots \int_{x_{N-1}}^L dx_N\, e^{-\beta\sum_{\alpha=1}^N \varphi_{i_\alpha,i_{\alpha+1}}(x_{\alpha+1}-x_\alpha)},
\eal
where, as a generalization of Eq.~\eqref{varphi_ij}, we have called
\beq
\label{varphi_ij_2}
\varphi_{ij}(x)=\varphi\left(\sqrt{x^2+(y_i-y_j)^2}\right).
\eeq

Now we consider an $M$-component 1D mixture where particles of species $i$ and $j$ interact via the pair potential given by Eq.~\eqref{varphi_ij_2}.
The corresponding grand partition function is
\bal
\label{Xi1D}
\Xi^{\text{1D}}(\beta,L,\{\mu_i\})=&1+\sum_{N=1}^\infty \sum_{i_1=1}^M\sum_{i_2=1}^M\cdots \sum_{i_N=1}^M \nn
&\times\frac{ e^{\beta\sum_{i=1}^M\mu_i N_i}}{\Lambda_{\mathrm{dB}}^{N}} \mathcal{Q}_{N,\{i_\alpha\}}^{\text{1D}}(\beta,L),
\eal
where $\mu_i$ and $N_i$ are the chemical potential and the number of particles of species $i$, respectively, and
\bal
\label{Q_N_1}
\mathcal{Q}_{N,\{i_\alpha\}}^{\text{1D}}(\beta,L)=&\int_0^L dx_1\int_{x_1}^L dx_2 \ldots \int_{x_{N-1}}^L dx_N\nn
&\times e^{-\beta\sum_{\alpha=1}^N \varphi_{i_\alpha,i_{\alpha+1}}(x_{\alpha+1}-x_\alpha)}.
\eal
Next, we assume that the reservoir in contact with the 1D system has the same chemical potential for all the species, i.e., $\mu_i=\mu$. In that case, the combination of Eqs.~\eqref{Xiq1D} and \eqref{Q_N_2} is equivalent to the combination of Eqs.~\eqref{Xi1D} and \eqref{Q_N_1}, except for the irrelevant term $(\delta y/\Lambda_{\mathrm{dB}})^N$
\footnote{Note that, in the thermodynamic limit, $\ln \Xi^{\text{q1D}}=\ln \Xi^{\text{1D}}+\langle N\rangle \ln (\delta y/\Lambda_{\mathrm{dB}})$, where $\langle N\rangle$ is the macroscopic number of particles.}.
Of course, this equivalence is preserved in the continuum limit $M\to\infty$.

In summary:
\begin{enumerate}[label=(\roman*)]
  \item
The transverse position, $y$, and the transverse distribution, $\phi^2(y)$, in the original q1D system correspond to the dispersity parameter and the associated mole fraction, respectively, in the 1D polydisperse system.
\item
The 1D interaction potential between two particles of species $y$ and $y'$,  $\varphi_{yy'}(x)$, is directly related to the interaction potential of the 2D system, $\varphi(r)$, as $\varphi_{yy'}(x)=\varphi(r)$ with $r=\sqrt{x^2+(y-y')^2}$.
\item
If the hard-core diameter of the 2D particles is denoted as $r=1$, meaning that the interaction potential $\varphi(r)$ becomes infinite for $r<1$, then the minimum separation between 1D particles of species $y$ and $y'$ can be expressed as $a_{yy'}=\sqrt{1-(y-y')^2}$.

\item
The 1D system is considered \emph{nonadditive} because $a_{yy'}\neq\frac{1}{2}\left(a_{yy}+a_{y'y'}\right)$. This contrasts with the approximate additive mixture considered in Ref.~\cite{PK92}.

\item
Since the transverse coordinates of particles in the original q1D system are not fixed, the species identities in the equivalent 1D system are also not fixed. This necessitates the condition of equal chemical potential in the 1D system. Therefore, we utilize the grand canonical ensemble in this Appendix as the simplest way to enforce this common chemical potential requirement. However, it is important to note that the equivalence holds with any other ensemble in the thermodynamic limit, $N\to\infty$, $L\to\infty$, with $\lambda=N/L=\text{const}$.
\end{enumerate}

Regarding the latter point, the exact properties of 1D systems are most effectively derived within the isothermal-isobaric ensemble framework. In this context, the probability distribution function for the first neighbor of a particle of species $i$ to be located at a distance $x$ and belonging to species $j$ is given by $P_{ij}^{(1)}\propto e^{-\bp}e^{-\beta\varphi_{ij}(x)}$ \cite{S16}. The $\ell$th-neighbor distribution, $P_{ij}^{(\ell)}(x)$, can be obtained by successive convolutions of $P_{ij}^{(1)}(x)$. Consequently, the Laplace transform, $\widetilde{P}_{ij}^{(\ell)}(s)$, of $P_{ij}^{(\ell)}(x)$ is expressed as the $\ell$th power of the matrix $\widetilde{P}_{ij}^{(1)}(s)\propto \Omega_{ij}(s+\bp)$. Finally, using $\lambda \phi_j^2 g_{ij}(x)=\sum_{\ell=1}^\infty P_{ij}^{(\ell)}(x)$, one obtains Eq.~\eqref{eq:rdf} in Laplace space.

\section{RDF in real space}\label{app:rdfrealspace}
By formally expanding in powers of $A^2$,  Eq.~\eqref{eq:rdf} can be rewritten as
\beq
\label{3.5e}
\widetilde{G}_{ij}(s)=\frac{A^2}{\lambda \phi_i\phi_j}\sum_{n=1}^\infty A^{2(n-1)} \left[\mathsf{\Omega}^n(s+\beta p)\right]_{ij}.
\eeq
Equation~\eqref{3.5e} implies that
\beq
\label{3.5ee}
{g}_{ij}(x)=\frac{A^2}{\lambda \phi_i\phi_j}\sum_{n=1}^{\lfloor{x/a(\epsilon)}\rfloor} A^{2(n-1)} \gamma_{ij}^{(n)}(x),
\eeq
where the function $\gamma_{ij}^{(n)}(x)$ denotes the inverse Laplace transform of $\left[\mathsf{\Omega}^n(s+\beta p)\right]_{ij}$. As will be shown later, $\gamma_{ij}^{(n)}(x)=0$ when $x\leq n a(\epsilon)$, providing justification for the inclusion of the floor function $\lfloor{x/a(\epsilon)}\rfloor$ in the upper limit of the summation in Eq.~\eqref{3.5ee}.

From Eq.~\eqref{eq:omega}, note first that
\begin{equation}
\Omega_{ij}(s+\beta p)=e^{\beta^*}\left[\widetilde{R}^{(1)}(s;a_{ij})-\nu\widetilde{R}^{(1)}(s;b_{ij})\right],
\end{equation}
where
\begin{equation}
\label{tRn}
\widetilde{R}^{(n)}(s;\alpha)\equiv \frac{e^{-(s+\beta p)\alpha}}{(s+\beta p)^n},\quad \nu\equiv 1-e^{-\beta^*}.
\end{equation}
The inverse Laplace transform of $\widetilde{R}^{(n)}(s;\alpha)$ is
\begin{equation}
\label{Rnx}
R^{(n)}(x;\alpha)=\frac{e^{-\beta p x}}{(n-1)!}(x-\alpha)^{n-1}\Theta(x-\alpha).
\end{equation}

The matrices $\widetilde{R}^{(1)}(s;a_{ij})$ and $\widetilde{R}^{(1)}(s;b_{ij})$ do not commute. As a consequence, the expansion of $\mathsf{\Omega}^n(s+\beta p)$ generates $2^n$ independent terms  involving the function $\widetilde{R}^{(n)}(s;\alpha)$. In particular,
\begin{widetext}
\begin{subequations}
		\begin{align}
		\left[\mathsf{\Omega}^2(s+\beta p)\right]_{ij}=&e^{2\beta^*}\sum_{k}\left[\widetilde{R}^{(2)}(s;a_{ik}+a_{kj})-\nu \widetilde{R}^{(2)}(s;a_{ik}+b_{kj})
		-\nu \widetilde{R}^{(2)}(s;b_{ik}+a_{kj})+\nu^2 \widetilde{R}^{(2)}(s;b_{ik}+b_{kj})\right],
	\end{align}
		\begin{align}
		\left[\mathsf{\Omega}^3(s+\beta p)\right]_{ij}=&e^{3\beta^*}\sum_{k_1,k_2}\left[\widetilde{R}^{(3)}(s;a_{i{k_1}}+a_{{k_1}{k_2}}+a_{{k_2} j})-\nu \widetilde{R}^{(3)}(s;a_{i{k_1}}+a_{{k_1}{k_2}}+b_{{k_2} j})
		-\nu \widetilde{R}^{(3)}(s;a_{i{k_1}}+b_{{k_1}{k_2}}+a_{{k_2}j})\right.\nn
		&-\nu \widetilde{R}^{(3)}(s;b_{i{k_1}}+a_{{k_1}{k_2}}+a_{{k_2} j})
		+\nu^2 \widetilde{R}^{(3)}(s;a_{i{k_1}}+b_{{k_1}{k_2}}+b_{{k_2} j})+\nu^2 \widetilde{R}^{(3)}(s;b_{i{k_1}}+a_{{k_1}{k_2}}+b_{{k_2} j})
		\nn&\left.
		+\nu^2 \widetilde{R}^{(3)}(s;b_{i{k_1}}+b_{{k_1}{k_2}}+a_{{k_2} j})-\nu^3 \widetilde{R}^{(3)}(s;b_{i{k_1}}+b_{{k_1}{k_2}}+b_{{k_2} j})\right].
	\end{align}
	\end{subequations}
\end{widetext}
Consequently, in real space,
\begin{subequations}
\beq
\gamma_{ij}^{(1)}(x)=e^{\beta^*}\left[R^{(1)}(x;a_{ij})-\nu R^{(1)}(x;b_{ij})\right],
\eeq
\bal
\gamma_{ij}^{(2)}(x)=&e^{2\beta^*}\sum_{k}\left[R^{(2)}(x;a_{ik}+a_{kj})-\nu R^{(2)}(x;a_{ik}+b_{kj})
	\right.\nn
	&\left.-\nu R^{(2)}(x;b_{ik}+a_{kj})+\nu^2 R^{(2)}(x;b_{ik}+b_{kj})\right],
\eal
\end{subequations}
and so on.

It is noteworthy that, for any pair $ij$, both $a_{ij}$ and $b_{ij}$ cannot be smaller than $a(\epsilon)$. Hence, all distinct functions of the form $R^{(n)}(x;\alpha)$ that contribute to $\gamma_{ij}^{(n)}(x)$ satisfy $\alpha\geq na(\epsilon)$. As a consequence, the presence of the Heaviside function in Eq.~\eqref{Rnx} establishes that $\gamma_{ij}^{(n)}(x)=0$ when $x\leq na(\epsilon)$, as anticipated earlier.
In particular, only $\gamma_{ij}^{(1)}(x)$ is needed in Eq.~\eqref{3.5ee} if $x\leq 2a(\epsilon)$, while only $\gamma_{ij}^{(1)}(x)$ and $\gamma_{ij}^{(2)}(x)$ contribute if $x\leq 3a(\epsilon)$.

\section{Derivation of Eq.~\eqref{eq:Zfromgx}}\label{app:rdffromgx}
Consider a generic 2D potential $\varphi(r)$ with the constraints (i) $\varphi(r)=\infty$ if $r<1$ and (ii) $\varphi(r)=0$ if $r>r_0$. Then, the 1D potential defined by Eq.~\eqref{varphi_ij_2} fulfills (i) $\varphi_{ij}(x)=\infty$ if $x<a_{ij}$ and (ii) $\varphi_{ij}(x)=0$ if $x>b_{ij}$. The smallest value of the set $\{a_{ij}\}$ is $a(\epsilon)$, which corresponds to $|y_i-y_j|=\epsilon$. Analogously, the maximum value of the set $\{b_{ij}\}$ is $r_0$,  corresponding to $y_i=y_j$. To guarantee that interactions are restricted to nearest neighbors, one must have  $r_0<2a(\epsilon)$.

Under the above conditions, the Laplace transform defined by Eq.~\eqref{Om0} can be written as
\beq
\label{Om}
\Omega_{ij}(s)=\int_{a(\epsilon)}^{r_0} dx\, e^{-s x} e^{-\beta\varphi_{ij}(x)}+\frac{e^{-s r_0}}{s},
\eeq
whose derivative is
\beq
\label{Om'}
\partial_s\Omega_{ij}(s)=-\int_{a(\epsilon)}^{r_0} dx\,x e^{-s x} e^{-\beta\varphi_{ij}(x)}-\frac{e^{-s r_0}}{s}\left(r_0+ \frac{1}{s}\right).
\eeq

Our aim is to express the equation of state in terms of the integrals
\beq
\label{3}
I_n\equiv \lambda\int_{a(\epsilon)}^{r_0} dx\, x^n g(x),\quad n=0,1.
\eeq
To that end, note that, in the interval $a_{ij}<x<2a_{ij}$, only the first-neighbor distribution function contributes to the partial RDF $g_{ij}(x)$ \cite{S16,MS23b}, i.e., $g_{ij}(x)=(A^2/\lambda\phi_i\phi_j)e^{-\bp x}e^{-\beta\varphi_{ij}(x)}$. Therefore, the total RDF in the range $a(\epsilon)<x<2a(\epsilon)$ is
\beq
\label{2}
g(x)=\frac{A^2}{\lambda}\sum_{i,j} \phi_i\phi_j e^{-\bp x}e^{-\beta\varphi_{ij}(x)},\quad a(\epsilon)<x<2a(\epsilon).
\eeq
As a consequence,
\beq
\label{3b}
I_n={A^2}\sum_{i,j} \phi_i\phi_j \int_{a(\epsilon)}^{r_0} dx\, x^n e^{-\bp x}e^{-\beta\varphi_{ij}(x)}.
\eeq
From Eqs.~\eqref{Om} and \eqref{Om'} we have
\begin{subequations}
	\begin{align}
	I_0=&{A^2}\sum_{i,j} \phi_i\phi_j\left[\Omega_{ij}(\bp)-\frac{e^{-\bp r_0}}{\bp}\right]\nn
	=&1-{A^2}\frac{e^{-\bp r_0}}{\bp}\sum_{i,j} \phi_i\phi_j,
\end{align}
	\begin{align}
	I_1=&-{A^2}\sum_{i,j} \phi_i\phi_j\left[\frac{\partial\Omega_{ij}(\bp)}{\partial \bp}+\frac{e^{-\bp r_0}}{\bp}\left( r_0+\frac{1}{\bp}\right)\right]\nn
	=&\frac{1}{\lambda}-A^2\frac{e^{-\bp r_0}}{\bp}\left(r_0+\frac{1}{\bp}\right)\sum_{i,j} \phi_i\phi_j,
\end{align}
\end{subequations}
where in the second steps we have used Eqs.~\eqref{eq:eigenproblem} and \eqref{eq:eos}, respectively.
Eliminating ${A^2}({e^{-\bp r_0}}/{\bp})\sum_{i,j} \phi_i\phi_j$ between both equations, we get
\beq
\label{9}
I_1=\frac{1}{\lambda}-\left(r_0+\frac{1}{\bp}\right)(1-I_0).
\eeq
Finally, using $\bp =Z\lambda$, Eq.~\eqref{9} yields
\beq
\label{11}
Z=\frac{1-I_0}{1-\lambda\left[r_0(1-I_0)+I_1\right]},
\eeq
which is the same as Eq.~\eqref{eq:Zfromgx}.


    \bibliography{C:/AA_D/Dropbox/Mis_Dropcumentos/bib_files/liquid}


\end{document}